\documentclass[reprint,superscriptaddress,aps,prb,floatfix]{revtex4-2}

\usepackage{amsmath,amssymb,amsfonts}
\usepackage{graphicx}
\usepackage{dcolumn}
\usepackage{bm}
\usepackage{color}
\usepackage{comment}
\usepackage[utf8]{inputenc}
\usepackage[english]{babel}
\usepackage{dsfont}
\usepackage{appendix}

\newcommand{\DU}{D_\text{\tiny U}}

\newcommand{\pRSB}{p_\text{\tiny RSB}}
\newcommand{\qRSB}{q_\text{\tiny RSB}}

\newcommand{\beq}{\begin{equation}}
\newcommand{\eeq}{\end{equation}}
\newcommand{\beqa}{\begin{eqnarray}}
\newcommand{\eeqa}{\end{eqnarray}}

\newcommand{\<}{\langle}
\renewcommand{\>}{\rangle}

\def\<{\langle}
\def\>{\rangle}

\newcommand{\dd}[1]{\mathrm{d}#1}

\DeclareMathOperator{\arctanh}{arctanh}

\begin{document}

\title{The Ising Spin Glass on Random Graphs at zero temperature:\\ Not all Spins are Glassy in the Glassy Phase}

\author{Gianmarco Perrupato}
\affiliation{Dipartimento di Fisica, Sapienza Universit\`a di Roma, P.le A. Moro 5, 00185 Rome, Italy}

\author{Maria Chiara Angelini}
\affiliation{Dipartimento di Fisica, Sapienza Universit\`a di Roma, P.le A. Moro 5, 00185 Rome, Italy}
\affiliation{Istituto Nazionale di Fisica Nucleare, Sezione di Roma I, P.le A. Moro 5, 00185 Rome, Italy}

\author{Giorgio Parisi}
\affiliation{Dipartimento di Fisica, Sapienza Universit\`a di Roma, P.le A. Moro 5, 00185 Rome, Italy}
\affiliation{Istituto Nazionale di Fisica Nucleare, Sezione di Roma I, P.le A. Moro 5, 00185 Rome, Italy}
\affiliation{Institute of Nanotechnology (NANOTEC) - CNR, Rome unit, P.le A. Moro 5, 00185 Rome, Italy}

\author{Federico Ricci-Tersenghi}
\affiliation{Dipartimento di Fisica, Sapienza Universit\`a di Roma, P.le A. Moro 5, 00185 Rome, Italy}
\affiliation{Istituto Nazionale di Fisica Nucleare, Sezione di Roma I, P.le A. Moro 5, 00185 Rome, Italy}
\affiliation{Institute of Nanotechnology (NANOTEC) - CNR, Rome unit, P.le A. Moro 5, 00185 Rome, Italy}

\author{Tommaso Rizzo}
\affiliation{Institute of Complex Systems (ISC) - CNR, Rome unit, P.le A. Moro 5, 00185 Rome, Italy}
\affiliation{Dipartimento di Fisica, Sapienza Universit\`a di Roma, P.le A. Moro 5, 00185 Rome, Italy}


\begin{abstract}
We investigate the replica symmetry broken (RSB) phase of spin glass (SG) models in a random field defined on Bethe lattices at zero temperature.
From the properties of the RSB solution we deduce a closed equation for the extreme values of the cavity fields.
This equation turns out not to depend on the parameters defining the RSB, and it predicts that the spontaneous RSB does not take place homogeneously on the whole system.
Indeed, there exist spins having the same effective local field in all local ground states, exactly as in the replica symmetric (RS) phase, while the spontaneous RSB manifests only on the remaining spins, whose fraction vanishes at criticality.
The characterization in terms of spins having fixed or fluctuating local fields can be extended also to the random field Ising model (RFIM), in which case the fluctuating spins are the only responsible for the spontaneous magnetization in the ferromagnetic phase.
Close to criticality we are able to connect the statistics of the local fields acting on the spins in the RSB phase with the correlation functions measured in the paramagnetic phase.
Identifying the two types of spins on given instances of SG and RFIM, we show that they participate very differently to avalanches produced by flipping a single spin.
From the scaling of the number of spins inducing RSB effects close to the critical point and using the $M$-layer expansion we estimate the upper critical dimension $\DU \ge 8$ for SG.
\end{abstract}

\maketitle

\section{Introduction}

Despite the simplicity of their microscopic definition, spin glasses (SGs) display so outstandingly complicated behaviors to become the benchmark for complex systems, and the inspiration of a vast literature of models. In general their Hamiltonian can be written as follows:
\begin{equation}
\label{eq:Hamiltonian}
\mathcal{H}(\underline{\sigma})=-\sum_{(ij)\in\mathcal{E}}J_{ij}\sigma_i\sigma_j-\sum_{i\in\mathcal{V}}H_{i}\sigma_i\;,
\end{equation}
where $\underline{\sigma}=(\sigma_1,\dots,\sigma_N)\in \{-1,1\}^N$ is a spin configuration of the system, $\mathcal{V}$ and $\mathcal{E}$ are respectively the vertex set and the edge set of a graph $\mathcal{G}=(\mathcal{V},\mathcal{E})$, and the $J_{ij}$'s and the $H_i$'s are random independent variables. A paradigmatic example is the mean field theory obtained on the fully connected graph, namely the so called Sherrington–Kirkpatrick (SK) model \cite{SK}. The solution of the SK model requires to introduce $n$ independent replicas of the system, that, after the average over disorder, interact according to an effective Hamiltonian that is symmetric under replica permutations. What turns out is that for small values of the temperature and external field there is a region in which the replica symmetry is spontaneously broken \cite{Parisi1979, parisi1980order, parisi1980sequence}. The critical line in the temperature-field plane separating the replica-symmetric (RS) phase from the replica-symmetry-broken (RSB) phase is called de Almeida-Thouless (dAT) line \cite{DeAlmeida1978}. The same mechanism is conjectured also to rule SGs on Bethe lattices (BL) \cite{Mezard2001,Mezard2003,panchenko2016structure,parisi2017marginally,concetti2019properties,de2018computation}, i.e.\ finite-connectivity random graphs in which the neighbourhood of a site taken at random is typically a tree up to a distance that is diverging in the thermodynamic limit. Exploiting the local tree-like structure of the graph, it is possible to use an iterative technique called cavity method that allows to write self-consistent equations for the order parameter at any finite step $k$ of RSB \cite{Mezard2001,Mezard2003}. However such equations are in general extremely complicated to solve, indeed the finite connectivity implies that the distribution of the local cavity fields, even within one pure state, cannot be easily parameterized. For this reason most of the studies of the RSB solution on the BL rely on expansions for large connectivities \cite{goldschmidt1990finite,de1989replica,goldschmidt1990replica,parisi2002spin,boschi2020free}, expansions near the critical line \cite{mottishaw1987replica,parisi2013critical,rizzo2013replica}, or on the analysis of the 1RSB ansatz of the cavity equation \cite{Mezard2001,Mezard2003}. It is worth underlining that already for $k=1$ the order parameter is a complicated object, namely a probability distribution of probability distributions, and the cavity equations are usually solved numerically, or by means of variational approximations. 

In this paper, we present a study of some properties of the exact RSB solution, focusing on the case of zero temperature. The region at $T=0$ is particularly interesting because, contrarily to the SK model in which the dAT line diverges, on the BL there is a transition in the external field \cite{parisi2014diluted}. A first question to ask is whether this point has the same critical properties exhibited at $T>0$. At finite temperature one can show that deviations from the replica-symmetric phase can be described by an effective theory for the standard replicated order parameter $q_{ab}$ \cite{parisi2013critical,rizzo2013replica} with coefficients that are expressed  in terms of the replica-symmetric solution at the critical point \cite{parisi2014diluted}. This allows to show that the critical properties in this case are exactly the same of the SK model. Surprisingly, as we are going to present here, this is not true at $T=0$ due to the emergence of a different physics.

The paper is organized as follows. In section 
\ref{sec:RSBT0} we introduce the extremes of the cavity field, and the distinction between spins having fixed or fluctuating local fields. The reader is referred to appendix \ref{sec:kRSBansatz} for a discussion about RSB at $T=0$, and to appendix \ref{sec:DistrExtrApp} for the derivation of the self-consistent equation for the probability distribution of the extremes. In section \ref{sec:RFIM} we show that also for the random field Ising model (RFIM) it is possible to carry out an analogous distinction between fixed and fluctuating sites, and that they can be studied with the same formalism of the extremes found for the SG problem. In section \ref{sec:criticalBehaviour} we study the critical behavior of the equation for the extremes, leaving the derivations to appendices \ref{sec:ExpansionCloseToCrit}, \ref{RFIMcritical} and \ref{sec:CompQeps}. In section \ref{sec:ConnectionToCorrFunct} we show that the properties of the extremes close to the critical point are the same of the zero-temperature correlations in the paramagnetic phase, leaving the derivations to appendix \ref{sec:LineParamagnPhase}. In section \ref{sec:avalanches} we discuss the relation between extremes and spin avalanches, both in the SG problem and in the RFIM. In section \ref{sec:FiniteDim} and appendix \ref{sec:GinzSGphase} we derive some consequences in finite dimension of the results obtained in the previous sections. In section \ref{sec:Conclusions} we present the conclusions of our study.

\section{Replica Symmetry Breaking at $T=0$ and cavity field extremes}
\label{sec:RSBT0}

The breaking of the replica symmetry implies the presence of many local ground states (LGS), that at $T=0$ correspond to configurations whose energy cannot be decreased by flipping any finite number of spins \cite{Mezard2003}. The lowest LGSs differ from the global GS by an extensive number of spins, but have energy differences of order one. 
As a consequence the local properties of the system depend on the LGS.
An essential prediction of the cavity method is that each LGS is in correspondence with a fixed point
of the RS cavity equations, also known as \emph{belief propagation} (BP) equations \cite{mezard2009information}:
\begin{eqnarray}
\label{eq:RScavEQ}
u_{i\to j}&=&\hat{u}_{J_{ij}}\bigg(H_i+\sum_{k\in\partial i\setminus j}u_{k\to i}\bigg)\;,\quad \forall (i\!\to\! j)\in\mathcal{D} \\
\label{eq:filterFunctU}
\hat{u}_J(h)&=&\text{sgn}(J h)\,\min\left\{|J|,|h|\right\}\;, 
\end{eqnarray}
where $\partial i\setminus j$ represents the set of neighbors of $i$ except $j$ and $\mathcal{D}$ is the set of \emph{directed} edges induced by the graph edge set $\mathcal{E}$.
The so-called cavity field $u_{i\to j}$ is the effective field induced on spin $j$ by spin $i$ along the directed edge $(i\!\to\! j)$ of the graph. A fixed point $\{u_{i\to j}\}_{(i\to j)\in\mathcal{D}}$ of Eq.~(\ref{eq:RScavEQ}) determines the global and local properties of the system in a specific LGS \cite{mezard2009information}. 
As an example we consider the magnetization $m_i^{\alpha}$ of the spin $i$ on the LGS $\alpha$, that is given by
\begin{equation}
\label{eq:magnetiz}
m_i^{\alpha}=\text{sgn}\big(h_i^{\alpha}\big),\quad h_i^{\alpha}=H_i+\sum_{j\in\partial i}u_{j\to i}^{\alpha}\;.
\end{equation}
At finite temperature the second-order nature of the transition is reflected by the fact that a given local value of the cavity field in the paramagnetic (PM) phase is replaced, just below the dAT line, by a local distribution (population) of fields centered around that value, with a width that grows continuously from zero departing from the dAT line. Also in this case, the elements of these populations correspond to the values taken by the cavity fields in the different fixed points of the finite temperature RS recursion relation \cite{Mezard2001}.

However at $T=0$ we find an essential difference with respect to the finite temperature case: only a {\it finite} fraction of the sites (that we call the RSB cluster) displays a non-zero width for the population of fields $h_i^\alpha$ (we say that the population of fields \emph{opens} up on these sites).
Sites not belonging to the RSB cluster have the same effective local field $h_i^\alpha$ on \emph{all} LGSs (that is the population of fields is \emph{closed} on a single value).

This feature emerges naturally from the RSB cavity equations (see appendices \ref{sec:kRSBansatz} and \ref{sec:DistrExtrApp}): while at finite temperature the fields associated with all the spins open and are promoted to populations of non-zero width upon crossing the dAT line, at zero temperature only the fields associated with a tiny fraction of the spins open up. Therefore in the SG phase we can distinguish between spins with closed populations of fields, whose support concentrates on a single value (as in the PM phase), and spins with open populations of fields that can take more than one value.

The simplest way to quantify this phenomenon is by looking at the \emph{extreme values} $u^{+},u^{-}$ taken by the cavity field on a generic directed edge of the graph
\begin{equation}
\label{eq:extrDef}
u^{+}=\max_{\alpha\in\text{LGS}}u^{\alpha},\quad u^{-}=\min_{\alpha\in\text{LGS}}u^{\alpha}.
\end{equation}
With this notation a site $i$ is closed if for all its neighbours $u^+_{j\to i}=u^-_{j\to i}$. While the complete characterization of the statistical properties of the local field requires the knowledge of the whole RSB order parameter (see appendix \ref{sec:kRSBansatz}), we find that the RSB cavity equation can be closed {\it exactly} on the following equation for the extreme values (see appendix \ref{sec:DistrExtrApp}):
\begin{equation}
\label{eq:eqExtremes1}
\begin{split}
u^\pm_{i\to j} &= f^{(\pm)}_{J_{ij}}(h^{+}_{i\to j},h^{-}_{i\to j})\;,\\
h^\pm_{i\to j} &= H_i+\sum_{k\in\partial i\setminus j}u^{\pm}_{k\to i}\;,
\end{split}
\end{equation}
where we have introduced the ordering functions
\begin{equation}
\label{eq:eqExtremes3}
\begin{split}
f^{(+)}_{J}(h^{+},h^{-})&=\max{\left\{\hat{u}_J(h^+),\hat{u}_J(h^-)\right\}}\;,\\
f^{(-)}_{J}(h^{+},h^{-})&=\min{\left\{\hat{u}_J(h^+),\hat{u}_J(h^-)\right\}}\;,
\end{split}
\end{equation}
and $\hat{u}_J(h)$ is defined by Eq.~(\ref{eq:filterFunctU}).
A first observation about the recursive equations \eqref{eq:eqExtremes1} is that they do not depend on the number of RSB steps and on the corresponding RSB parameters. This is somehow consistent with the fact that a variation of the RSB parameters corresponds to exploring different LGS \cite{mezard1987spin} and should
not have any influence on sites that are not on the RSB cluster.
Eqs.~\eqref{eq:eqExtremes1} can be solved for a given instance of the disorder, or in the distributional sense, in order to determine the joint probability distribution of the couple $(u^+,u^-)$. By introducing the median $u$ and the width $\Delta$,
\begin{equation}
u^+=u+\Delta/2,\quad u^-=u-\Delta/2\, ,
\end{equation}
the problem can be equivalently rewritten also in terms of a distribution $Q(u,\Delta)$. In the PM phase we have $Q(u,\Delta)=Q_{RS}(u)\,\delta(\Delta)$, corresponding to all populations being closed, while in the SG phase there is a finite fraction $p$ of populations with $\Delta>0$. Therefore we can write the following decomposition:
\begin{equation}
\label{eq:decompOrdPar}
Q(u,\Delta)=p\,\mathcal{Q}_o(u,\Delta)+(1-p)\,\mathcal{Q}_c(u)\,\delta(\Delta),
\end{equation}
where $\mathcal{Q}_o$ and $\mathcal{Q}_c$ are the distributions of the extremes conditioned, respectively, to the open and closed populations. As anticipated before we found that at $T=0$ in the RSB phase $p<1$ strictly.

\begin{figure}[t]
\includegraphics[width=\columnwidth]{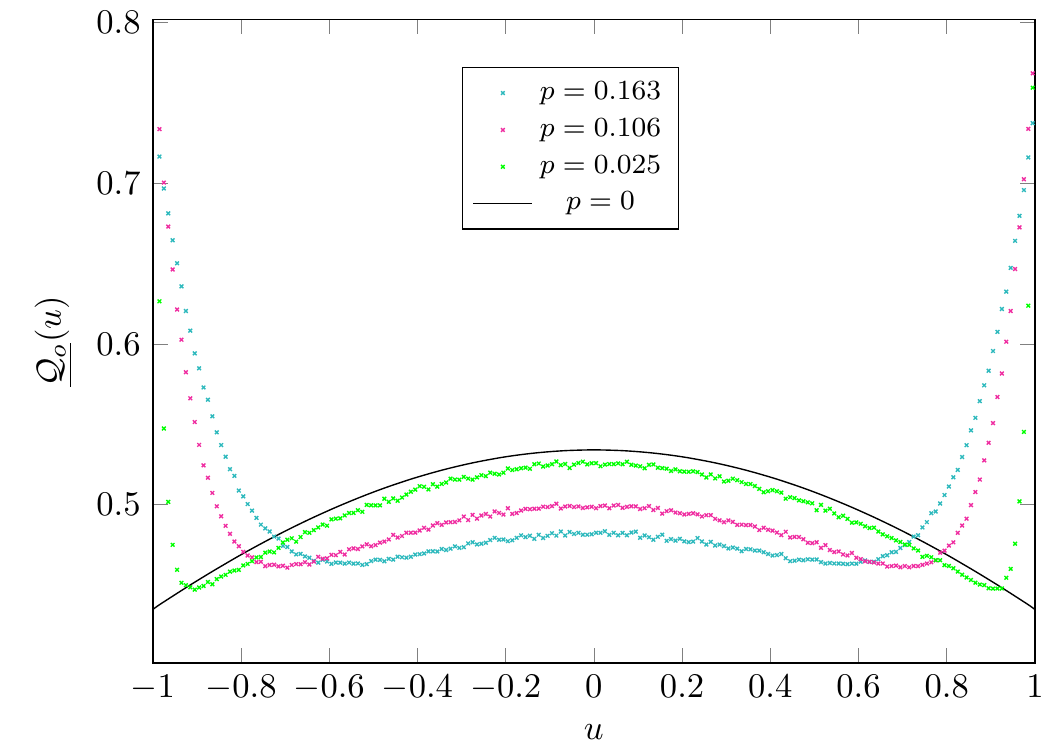}\\
\vspace{5mm}
\includegraphics[width=\columnwidth]{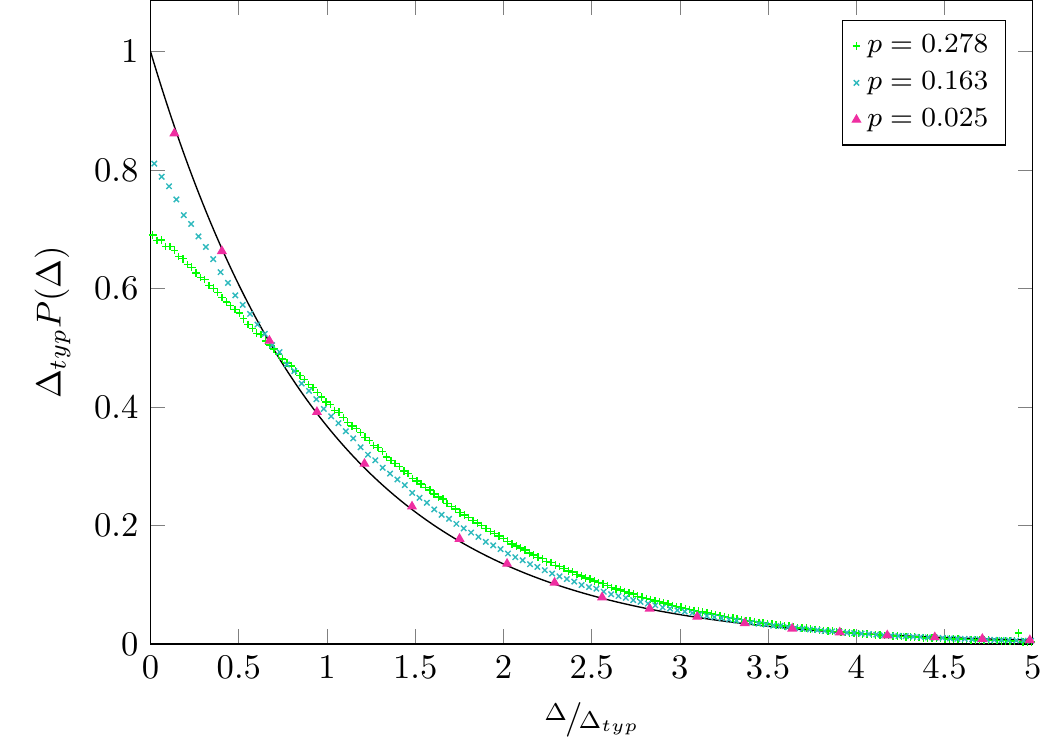}
\caption{\label{fig:limitDistr}
Probability distribution $\underline{\mathcal{Q}_o}(u)$ of the median value $u=(u^++u^-)/2$, conditioned to $\Delta:=u^+-u^->0$ (top panel) and probability distribution of the rescaled width of the open populations of fields (bottom panel), for different values of $p$, corresponding to different values of $\sigma_H$. The data are obtained on a Bethe lattice with $z=3$, $J_{ij}=\pm 1$ and $H_i\sim\mathcal{N}(0,\sigma_H)$ through a population dynamics algorithm with population size $N=10^7$. The continuous line in the top panel represents the theoretical estimate, $g^d(u)$, following Eq.~\eqref{eq:ordParamExp}, while the one in the bottom panel below is the exponential law $\exp(-x)$.
}
\end{figure}

It is important to note that, from a technical point of view, a fundamental ingredient for the existence of closed populations is that the function $\hat{u}_J(h)$ is constant for $|h|>|J|$. This means that it acts like a filter, closing an open couple $(h^+,h^{-})$ if $h^->|J|$ or $h^+<-|J|$.  This property does not follow from a particular choice of the distribution of the couplings and the external fields,  but it is a consequence of the structure of the zero temperature cavity equations. For this reason, even if for simplicity in the following we refer to the case of bimodal couplings $J_{ij}=\pm 1$ and Gaussian external fields $H_i\sim\mathcal{N}(0,\sigma_H)$, we expect this ``closure'' phenomenon to be more general. 

The order parameter $Q(u,\Delta)$ can be numerically found by means of a population dynamics algorithm \cite{mezard2009information,Mezard2003,Mezard2001,mezard2002random}. 
In the regime of small $p$ it is particularly convenient to solve the distributional equation for $Q(u,\Delta)$ by using two populations for representing separately $\mathcal{Q}_o$ and $\mathcal{Q}_c$. Indeed by using a unique vector of couples of size $N$, close to criticality one should set $N\propto p^{-1}$ in order to sample a constant number of open couples. Instead by separating the closed couples from the open ones, it is possible to work at fixed resolution without changing $N$.

\begin{figure}[t]
\centering
\includegraphics[width=\columnwidth]{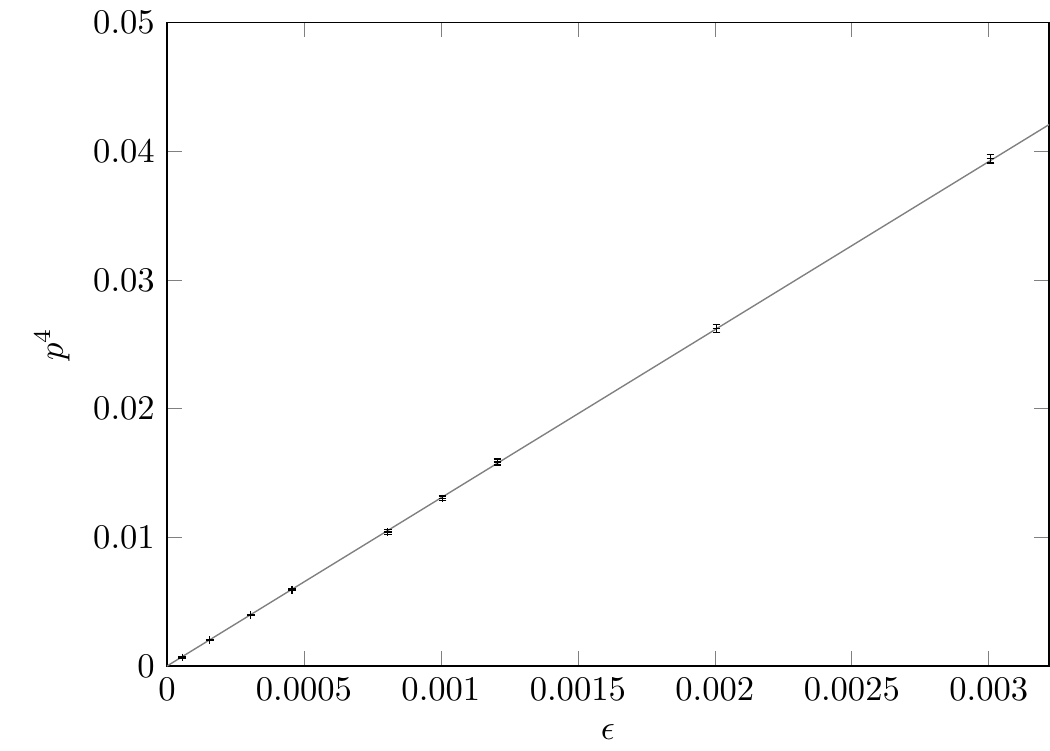}
\caption{\label{fig:epsilon}
Probability $p$ of drawing an open couple $(u^+,u^-)$ as a function of distance from the critical point $\epsilon=\hat\sigma_H-\sigma_H$. The data are obtained on a Bethe lattice with $z=3$, $J_{ij}=\pm 1$ and $H_i\sim\mathcal{N}(0,\sigma_H)$ through a population dynamics algorithm with population size $N=10^8$. The continuous line represents the analytical prediction close to the critical point $p\approx\kappa\epsilon^{1/4}$, with $\kappa\approx 1.90$ (see appendix \ref{sec:CompQeps}).
}
\end{figure}

In Fig.~\ref{fig:limitDistr} we show the marginal distributions
\begin{equation*}
\underline{\mathcal{Q}_o}(u)=\int\mathrm{d}\Delta\, \mathcal{Q}_o(u,\Delta)\;,\quad P(\Delta)=\int\mathrm{d}u\, \mathcal{Q}_o(u,\Delta)\;,
\end{equation*}
computed on a BL with $z=3$ through a population dynamics algorithm with population size $N=10^7$. Results are shown for different values of $p$ that correspond to different values of $\sigma_H$ as we are going to explain. At $T=0$ the model is paramagnetic for $\sigma_H>\hat\sigma_H$ and in the SG phase for $\sigma_H<\hat\sigma_H$. We call $\epsilon=\hat\sigma_H-\sigma_H$ the distance from the critical point. In Fig.~\ref{fig:epsilon} we show numerical data supporting the scaling $p\sim\kappa\epsilon^{1/4}$, that will be discussed below (see Sec.~\ref{sec:criticalBehaviour}).

\section{Random Field Ising Model}
\label{sec:RFIM}

Interestingly enough, Eqs.~\eqref{eq:eqExtremes1} admit a non trivial solution also if we set all the couplings equal to a constant, $J_{ij}=J>0$, while keeping the disorder only in the external fields, $H_i\sim\mathcal{N}(0,\sigma_H)$. This case corresponds to another prototypical disordered system, the Random Field Ising Model (RFIM). The RFIM undergoes a ferromagnetic transition in the $T-\sigma_H$ plane \cite{morone2014large}. The ferromagnetic line, on which the ferromagnetic susceptibility diverges, coincides with the dAT line at $T=0$ \cite{parisi2014diluted}, and therefore in this case the critical point is the same for both problems. Despite the apparent similarity with the SG, for the RFIM it has been rigorously proven that the SG susceptibility is always upper-bounded by the ferromagnetic susceptibility \cite{krzakala2010elusive,chatterjee2015absence}. Consequently, there cannot be a SG phase out of the critical ferromagnetic line. 
Moreover, even if the thermodynamics of the RFIM is always RS, the free energy landscape close to the critical point is characterized by the presence of many metastable states, that on the BL are associated with the many solutions of the RS cavity equations \cite{perugini2018improved}.

When $J>0$, Eqs.~\eqref{eq:eqExtremes1} simplify to
\begin{equation}
\label{eq:ExtrvalRF}
(u^+_{i\to j},u^-_{i\to j})=\big(\hat{u}_{J}(h^{+}_{i\to j}),\hat{u}_{J}(h^{-}_{i\to j})\big).
\end{equation}
It is important to observe in Eq.~\eqref{eq:ExtrvalRF} that the ``$+$'' and ``$-$'' cavity fields satisfy separately the RS cavity equations, defining two actual fixed points of BP. Differently, in the general case of Eq.~\eqref{eq:eqExtremes1} the ``$+$'' and ``$-$'' fields are coupled and do not correspond to two fixed points of BP. In the RFIM the ``$+$'' and ``$-$'' fixed points are, respectively, those with maximum and minimum magnetizations \cite{perugini2018improved}, indeed we can write
\begin{equation}
\label{eq:magIneqRFIM}
m_i^-=\text{sign}(h_i^{-}) \leq \text{sign}(h_i^{\alpha}) \leq \text{sign}(h_i^{+})=m_i^{+}\;,
\end{equation}
where we denoted by $h^+$ and $h^-$ the extreme values of the total local field on a spin, see Eq.~\eqref{eq:magnetiz}.

In the following we will call \emph{frozen} a spin $i$ for which $m_i^+=m^-_i$, and \emph{unfrozen} otherwise. In general the fraction $n_\text{unf}$ of unfrozen spins is smaller than the fraction of spins with open populations. Indeed, due to the sign operation in Eq.~\eqref{eq:magIneqRFIM}, there may exist frozen spins having an open distribution of the total local field, i.e.\ $h^-<h^+$, but with the extreme fields of the same sign.

Note that if the external field $H_i$ is not symmetrically distributed, in the thermodynamic limit the fraction of open populations should vanish both in the ferromagnetic and in the PM phase due to the uniqueness of the thermodynamic state. Conversely, if the disorder is symmetric, in the ferromagnetic phase the distribution of the extremes $Q^\text{\tiny RFIM}(u,\Delta)$ is non-trivial, since there are two thermodynamic states with opposite global magnetization, that are associated with two distinct fixed points of the BP equations. By symmetry the unfrozen spins should be the only ones contributing to the average magnetization
\begin{multline}
\label{eq:NfrozEquiv}
m=\frac{1}{N}\sum_{i\in\mathcal{V}} m_i^+=-\frac{1}{N}\sum_{i\in\mathcal{V}} m_i^- =\frac{1}{2N}\sum_{i\in\mathcal{V}}(m_i^+-m_i^-)=\\
=\frac{1}{N}\sum_{i\in\mathcal{V}}\mathds{1}(m_i^+\neq m_i^-) = n_\text{unf} = 1-n_\text{fr}\;,
\end{multline}
where $n_\text{fr}$ is the fraction of frozen spins.

\section{Critical behavior}
\label{sec:criticalBehaviour}

We discuss now the critical behavior of the extremes. Two combined effects occur approaching the dAT line from the SG phase: {\it both the fraction of open populations and their width go to zero}. This allows us to linearize the equation for the extremes with respect to $\mathcal{Q}_o$ close to $\hat{\sigma}_H$, leading to the following asymptotic expression (see appendix \ref{sec:ExpansionCloseToCrit}):
\begin{equation}
\label{eq:ordParamExp}
\mathcal{Q}_o(u,\Delta) \approx g^d(u) \, \frac{1}{\Delta_{typ}}e^{-\Delta/\Delta_{typ}},\quad \Delta_{typ} \propto p,
\end{equation}
where $g^d(u)$ is the eigenvector associated with the maximum eigenvalue of the linearized operator (see Eq.~(26) in Ref.~\cite{parisi2014diluted} and appendix \ref{sec:ExpansionCloseToCrit}).
In Fig.~\ref{fig:limitDistr} we compare the analytical prediction \eqref{eq:ordParamExp} with the numerics and indeed, in the limit $p\to 0$, we find that $\underline{\mathcal{Q}_o}(u)$ approaches $g^d(u)$ and $P(\Delta)$ converges to the exponential distribution.

An expansion similar to the one leading to Eq.~\eqref{eq:ordParamExp} can be performed also for the RFIM equations \eqref{eq:ExtrvalRF}, by using the fact that the global magnetization $m$ is a small parameter in the proximity of $\hat{\sigma}_H$ (see appendix \ref{RFIMcritical}). This allows to show that the dependence of $m$ on the distance from the critical point $\epsilon=\hat{\sigma}_H-\sigma_H$ is that of a standard ferromagnetic mean-field transition $\epsilon \sim m^2$.

\begin{figure}[t]
\centering
\includegraphics[width=\columnwidth]{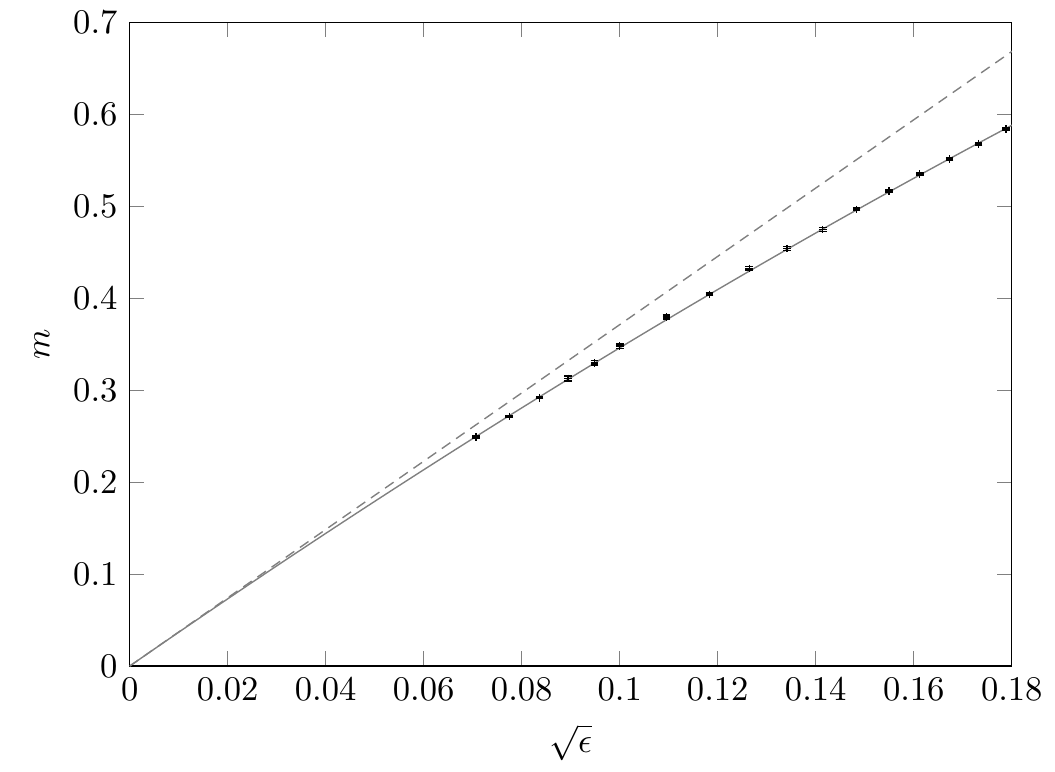}
\caption{\label{fig:magnSito}
Average magnetization $m$ in the RFIM as a function of the distance from the critical point $\epsilon$. The data are obtained on a Bethe lattice with $z=3$, $J=1$ and $H_i\sim\mathcal{N}(0,\sigma_H)$ through a population dynamics algorithm with population size $N=10^6$. The continuous line represents a quadratic fit $a\,\sqrt{\epsilon}+b\,\epsilon$, where $a=3.70(1)$ and $b=-2.43(6)$. The dashed line represents the analytical prediction close to the critical point $m\approx\alpha\sqrt{\epsilon}$, with $\alpha\approx 3.71$ (see appendix \ref{RFIMcritical}).
}
\end{figure}

It is interesting to note that if the disorder is symmetric, the fact that the order parameter is statistically symmetric implies that the joint distribution of the couples should be the same in both problems:
\begin{equation}
\label{eq:equivSimm}
Q^\text{\tiny SG}(u^+,u^-)= Q^\text{\tiny RFIM}(u^+,u^-)\equiv Q(u^+,u^-)\;.
\end{equation}
This implies a relation between $m$ and $p$. Indeed from Eq.~\eqref{eq:NfrozEquiv} and Eq.~\eqref{eq:equivSimm} we have
\begin{equation}
\label{eq:nUnfM}
m=n_\text{unf}^\text{\tiny RFIM}=n_\text{unf}^\text{\tiny SG}\;.
\end{equation}
At this point from the expansion leading to Eq.~\eqref{eq:ordParamExp}, we get $n_\text{unf}^\text{\tiny SG}\sim p^2$, and therefore (see appendix \ref{sec:CompQeps})
\begin{equation}
\label{eq:scalq}
\epsilon \sim m^2 \sim p^4\;.
\end{equation}
In Figs.~\ref{fig:epsilon} and \ref{fig:magnSito} we compare the analytical predictions in Eq.~\eqref{eq:scalq} with the numerics, obtaining a very accurate agreement.

\section{Connection to correlation functions}
\label{sec:ConnectionToCorrFunct}

\begin{figure}
\centering
\includegraphics[width=\columnwidth]{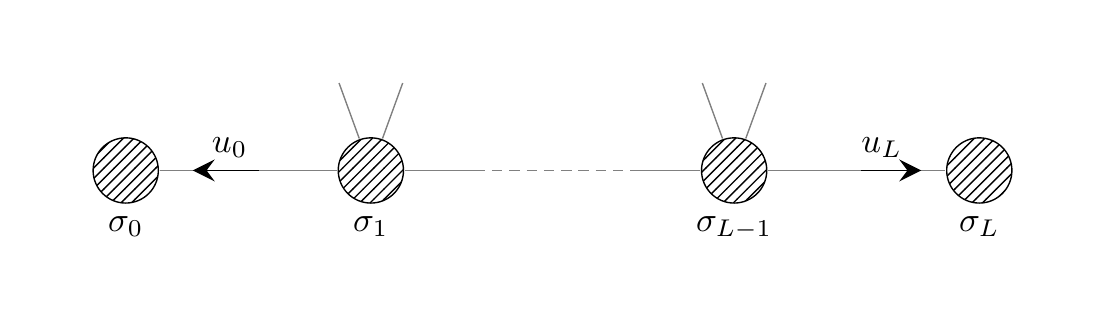}
\caption{Representation of a cavity chain of length $L$, in which all sites have connectivity $z=4$ but $\sigma_0$ and $\sigma_L$, that have one neighbour.\label{fig:ChainParamagnetic}} 
\end{figure}
We now want to show that there is a connection between the physics of the extremes in the RSB phase and that of the correlations in the PM phase.
The correlation between two spins $\sigma_0,\sigma_L$ at distance $L$ can be studied in the PM phase of both the SG and the RFIM by characterizing the properties of the effective two-spins Hamiltonian \cite{angelini2020loop,angelini2022unexpected}:
\begin{equation}
\label{eq:twoSpinHamilt}
\mathcal{H}_L(\sigma_0,\sigma_L)=-u_0 \sigma_0 - J_L \sigma_0 \sigma_L - u_L \sigma_L \;,
\end{equation}
where $J_L$ is an effective coupling and $u_0,u_L$ are effective fields (see Fig.~\ref{fig:ChainParamagnetic}).
It turns out that there is always a finite probability that $J_L$ is strictly equal to zero, implying a zero connected correlation function between $\sigma_0$ and $\sigma_L$. Furthermore, as discussed in Ref.~\cite{angelini2020loop,angelini2022unexpected}, there are two combined effects leading to a decrease of the connected correlation when $L$ tends to infinity: the fraction of non-zero effective couplings goes to zero (exponentially in $L$) {\it and} the average value of the non-zero couplings goes to zero as $1/L$. The analogy with the physics of the open populations in the SG phase is not fortuitous. Calling $u^+_L$ and $u^-_L$ respectively the maximum and the minimum fields acting on $\sigma_L$ when fixing $\sigma_0=\pm 1$, we have that
\begin{equation}
u^{\pm}_L = u_L \pm |J_L|\;.
\end{equation}
The iterative cavity equations for $u^{\pm}_L$ are the following (see appendix \ref{sec:LineParamagnPhase})
\begin{equation}
\label{eq:iterExtrchain1}
\begin{split}
(u_{L+1}^{+},u_{L+1}^{-})&\stackrel{d}{=}\big(f^{(+)}_{J}(h^{+}_L,h^{-}_L),f^{(-)}_{J}(h^{+}_L,h^{-}_L)\big),\\
h^{\pm}_L&\stackrel{d}{=}H+\sum_{i=1}^{z-1}u_i+u^{\pm}_L,
\end{split}
\end{equation}
where $f^{(\pm)}$ are the ones defined in Eq.~\eqref{eq:eqExtremes3} while the $u_i$'s are drawn from the RS cavity distribution. Eq.~\eqref{eq:iterExtrchain1} is formally analogous to Eq.~\eqref{eq:eqExtremes1} for the extremes in the case in which only one couple $(u^+_{k\to i},u^-_{k\to i})$ in Eq.~\eqref{eq:eqExtremes1} is open. This is the most likely case close to the critical point when $p\to 0$, and so the analogy holds at criticality.

In the large $L$ limit (see appendix \ref{sec:LineParamagnPhase} for the derivations), the joint distribution of non-zero effective couplings and effective fields is given by \cite{angelini2020loop,angelini2022unexpected}
\begin{equation}
\label{eq:distriJparamagnetic}
P^{(L)}(u_0,J_L,u_L) \approx L\, \lambda^L g^d(u_0)\, g^d(u_L)\, \frac{1}{2\,J_{typ}}\, e^{-|J_L|/J_{typ}}\;,
\end{equation}
where the eigenvalue $\lambda$ tends to $\lambda_c=1/(z-1)$ at the critical point, and
\begin{equation}
J_{typ}\sim\frac{1}{L}\;.
\end{equation}
Note that both the distributions of $J_L$ in the PM phase and that of $\Delta$ in the SG phase (see Eq.~\eqref{eq:ordParamExp}) are exponential. For $J_L$ this fact comes from the large $L$ limit and it holds for any distance from the critical point, while for the RSB populations this is true just close enough to the critical point.


\section{Avalanches}
\label{sec:avalanches}

Until now we have discussed the statistical properties of the extremes. However, Eq.~\eqref{eq:eqExtremes1} can be solved also for a specific instance of the disorder (see appendix \ref{sec:DistrExtrApp}), thus obtaining the information about which spins are frozen and which are unfrozen. On a given graph the presence of many LGSs (or metastable states) is connected with the phenomenon of non-linear responses to external perturbations. In particular, the addition of an $O(1)$ external local field on a site may result in a collective rearrangement (avalanche) of $O(N)$ spins \cite{le2009statistics,le2010avalanches,tarjus2013avalanches,angelini2020loop}.
Here we want to give evidence that both in the SG problem and in the RFIM the response to external perturbations is highly non homogeneous, involving typically the unfrozen spins much more that the frozen ones.

Consider the following numerical experiment. Given an instance of the problem we solve the equations for the extremes, that is Eq.~\eqref{eq:eqExtremes1} for SG or Eq.~\eqref{eq:ExtrvalRF} for the RFIM, to understand which spins are frozen and which are unfrozen. At the same time we solve by BP the RS cavity equations \eqref{eq:RScavEQ} and we call $\underline{\sigma}^{\alpha}$ the configuration of the spins obtained from the BP fixed point via Eq.~\eqref{eq:magnetiz}. Then we choose at random a spin $i$, we perturb locally the system by forcing the flip of that spin from $\sigma_i^{\alpha}$ to $-\sigma_i^{\alpha}$ and we solve again the BP equations \eqref{eq:RScavEQ}. In this way we obtain a new configuration $\underline{\sigma}^{\beta}$ that we compare with $\underline{\sigma}^{\alpha}$.  
We call avalanche the set of spins that change sign after the perturbation.

\begin{figure}[t]
\centering
\includegraphics[width=0.95\columnwidth]{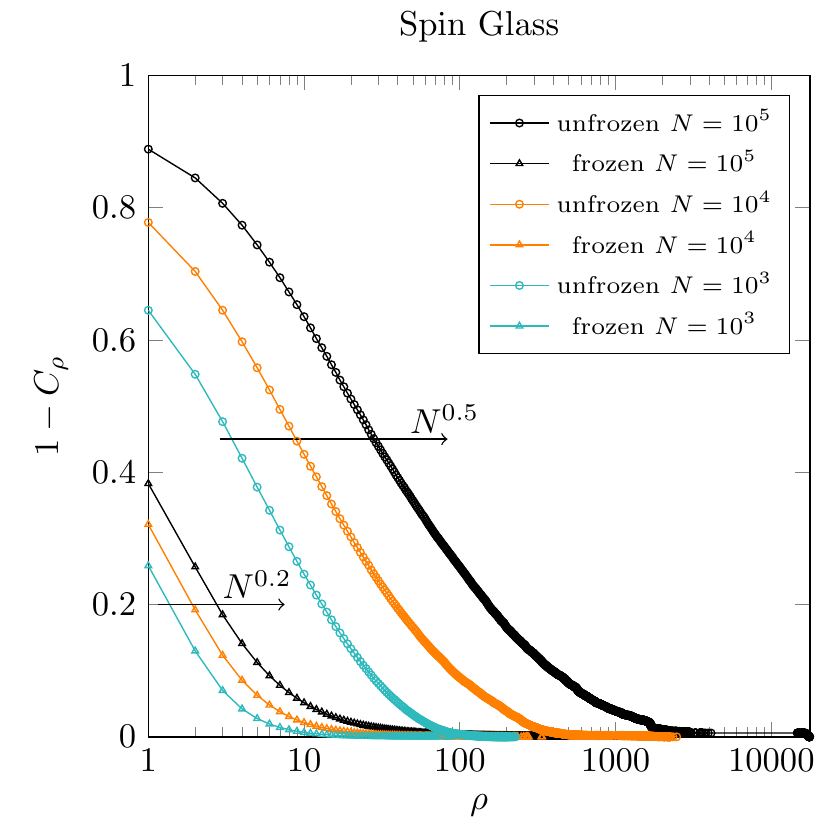}\\
\includegraphics[width=0.95\columnwidth]{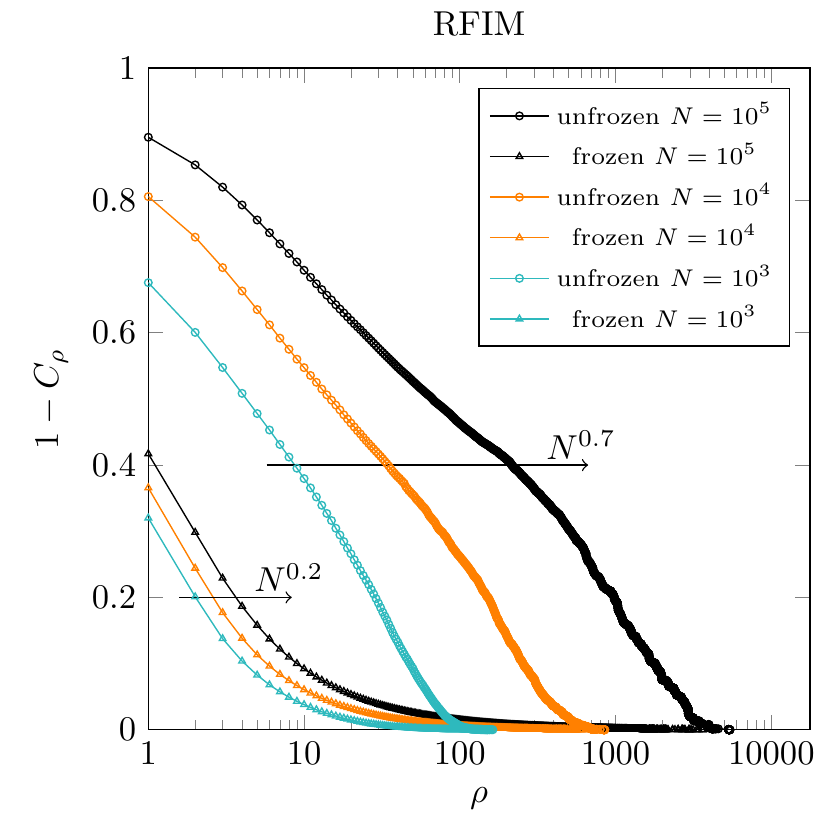}
\caption{Plot of the instance-averaged cumulative distribution of the participation in the Spin Glass problem (upper panel) and in the RFIM (lower panel). Data have been averaged over $10^6/N$ instances defined on a Bethe lattice of fixed degree $z=3$, $J_{ij}=\pm1$ (for SG) or $J_{ij}=1$ (for RFIM) and $H_i\sim\mathcal{N}(0,\sigma_H)$, with $\sigma_H\approx \hat\sigma_H$. Unfrozen spins participate much more to avalanches, and this is confirmed by the scaling of the typical values of $\rho$ shown by the arrows.
\label{fig:aval}
}
\end{figure}

For each instance of size $N$ we can generate $N$ different avalanches by perturbing different spins. We call $\rho_i$ the \emph{participation} of spin $i$, i.e.\ the number of avalanches (among the $N$ we generate) flipping it. For each instance $I$ we compute the empirical probability distribution of the participation $p_I(\rho)$ by computing the normalized histogram of the $\{\rho_i\}$ for that instance. We present the results in terms of the cumulative distribution of the participation $\rho$ averaged over the instances
\begin{equation}
C_{\rho} = \mathbb{E}_I \sum_{\rho'=0}^{\rho} p_I(\rho')\;.
\end{equation}
In Fig.~\ref{fig:aval} we show $1-C_{\rho}$ for the SG and the RFIM on a BL with fixed degree $z=3$ very close to the critical point $\hat{\sigma}_H=1.037$. In both plots we compare the cumulative distributions of the participation conditioning to frozen and unfrozen spins.
It is very evident that unfrozen spins participate much more to avalanches than frozen spins (please note the logarithmic scale on the $x$ axis).
In order to quantify this difference we have computed the scaling with $N$ of the typical values of the participation: while for frozen spins the typical participation scales like $N^{0.2}$, for unfrozen spins the typical participation scales with a much larger exponent ($N^{0.5}$ in the SG problem and $N^{0.7}$ in the RFIM).


\section{Consequences in finite dimensions}
\label{sec:FiniteDim}
Up to now we have discussed the properties of the RSB phase on a BL. Now we ask if and how this picture is modified in finite dimensions.
Recently a new loop expansion has been introduced in Ref.~\cite{altieri2017loop}, where the BL solution is the zero-th order term, and the effects of short loops are introduced perturbatively, in order to investigate the finite dimensional behavior of the system. In the following we will look at the first order term of this expansion checking whether the introduction of topological loops changes the BL picture described above.

A fundamental open problem in the statistical physics of disordered systems consists in determining if the SG problem on a $D$-dimensional lattice has a glassy transition in the presence of an external field. The so-called upper critical dimension $\DU$ is the dimension below which the fluctuations associated with the short-range interactions become so large that the MF picture does not predict the correct critical behavior anymore. The standard Renormalization Group (RG) approach, based on a field theoretical expansion around the MF fully-connected (FC) solution, leads to $\DU=6$. For $D \leq 6$ the coupling constants of the theory run away to infinity under the RG equations, implying the disappearance of the perturbative stable RG fixed point (FP) \cite{bray1980renormalisation,moore2011disappearance}. 
However a transition could still exist below $D=6$, and a possibility is that the associated FP cannot be reached continuously from the MF-FC one by lowering the dimension. Note that in the SK model there is no transition at $T=0$, implying that the expansion around the MF-FC FP is well defined only at finite temperature. Therefore, as suggested by some authors, a possibility is that there exists a relevant fixed point located at zero temperature \cite{parisi2012replica,angelini2015spin,angelini2017real,urbani2022field}. The expansion around the BL solution represents the perfect candidate for investigating this new FP, because on a BL the SG model exhibits a phase transition at zero temperature, in contrast to the FC theory. Such expansion has been applied to the SG with external field starting from the PM phase in Ref.~\cite{angelini2022unexpected}, leading to $\DU \geq 8$. Here we want to show that the Ginzburg criterion from the RSB phase leads to the same result. We refer the reader to appendix \ref{sec:GinzSGphase} for all the derivations of the results discussed in this section. 

The expansion around the BL has been made rigorous through the so-called $M$-layer construction: one starts from the model defined on an arbitrary graph $\mathcal{G}=(\mathcal{V},\mathcal{E})$, e.g.\ the $D$-dimensional hyper-cubic lattice. Then one replicates $M$ times $\mathcal{G}$ and, for each edge $(i,j)$ of $\mathcal{G}$, the $M$ copies of $\sigma_i$ and $\sigma_j$ are linked via a random permutation, that is each copy of $\sigma_i$ is linked to a randomly chosen copy of $\sigma_j$.
The ``rewired'' graph made of $M|\mathcal{V}|$ nodes and $M|\mathcal{E}|$ edges converges to a BL in the large $M$ limit. The observables in the original model ($M=1$) can be obtained by an expansion in powers of $1/M$.

In general if one gets so close to the BL critical point that the correlation length becomes comparable to the typical size of the tree-like neighbourhood of a node, one expect to see a deviation from the mean-field behavior. In particular if the model defined on $\mathcal{G}$ has a non-MF nature, one expects the $1/M$ corrections to diverge at the BL critical point. Vice versa, if the model defined on $\mathcal{G}$ has a MF nature, due to universality, the $M$-layer construction should not change its critical behavior.

Given a generic connected correlation $C(x)$ between two points at distance $x$ on $\mathcal{G}$ one finds \cite{altieri2017loop}:
\begin{equation}
\label{eq:leadingMcorr}
C(x)=\frac{1}{M}\sum_{L=1}^{\infty} \mathcal{N}_\text{\tiny NBP}(x,L)\, C^\text{\tiny BL}(L)\;,
\end{equation}
where $C^\text{\tiny BL}(L)$ is the same correlation computed between two points at distance $L$ on the BL and $\mathcal{N}_\text{\tiny NBP}(x,L)$ is the number of non-backtracking paths of length $L$ connecting the two points at distance $x$ on the original lattice \footnote{Note that Eq.~\eqref{eq:leadingMcorr} takes an analogous form to that of the finite size corrections to disorder models on sparse graphs \cite{ferrari2013finite,parisi2020random}}. For a $D$-dimensional hyper-cubic lattice the latter reads
\begin{equation}
\label{eq:combinTermMlayer}
\mathcal{N}_\text{\tiny NBP}(x,L) \propto (2D-1)^L \exp\left(-x^2/(4L)\right) L^{-D/2}\;.
\end{equation}
In the broken phase, the Ginzburg criterion consists in comparing the fluctuations of the order parameter on the correlation length scale, with the square of its average. In the RSB phase a convenient local order parameter is the RSB cluster indicator function, equal to 1 if a site is on the RSB cluster and 0 otherwise.
A site is on the RSB cluster if the distribution of its local field on the many LGS's is open and this happens if at least one of the cavity fields arriving from its $z$ neighbors is open. Thus the average order parameter is $\pRSB=1-(1-p)^z$.

The fluctuations of the order parameter are related to the probability $\qRSB(x)$ that two sites at distance $x$ on the lattice are both on the RSB cluster. The Ginzburg parameter $G(x)$ is therefore given by
\begin{equation}
\label{eq:defCorrx}
G(x)=C(x)/\pRSB^2,\quad C(x)=\qRSB(x)-\pRSB^2.
\end{equation}
From Eq.~\eqref{eq:leadingMcorr} we can write $C(x)$ in terms of the same quantity $C^\text{\tiny BL}(L)$ computed on the BL. Close to criticality, using the expansion in Eq.~\eqref{eq:ordParamExp}, we obtain the following expression for $C^\text{\tiny BL}$ (see appendix \ref{sec:GinzSGphase}):
\begin{equation}
\label{eq:FluctOrdParBLmain}
C^\text{\tiny BL}(L)=\qRSB^\text{\tiny BL}(L)-\pRSB^2\approx \pRSB^2\,L^3\lambda^L\;,
\end{equation}
where $\lambda(z-1)=1-a\pRSB$ with $a$ a constant, and $\qRSB^\text{\tiny BL}(L)$ is the probability that two sites at distance $L$ on the BL have both an open distribution of local fields. At this point, by using Eqs.~\eqref{eq:leadingMcorr}, \eqref{eq:combinTermMlayer} and \eqref{eq:defCorrx}, we can compute the Fourier transform $\widetilde{G}(q)$ of the Ginzburg parameter $G(x)$:
\begin{equation}
\widetilde{G}(q) \propto \frac{1}{M}(a\,\pRSB+q^2)^{-4}\;.
\end{equation}
Therefore in real space, if we rescale $x$ with the correlation length $\xi=O(\pRSB^{-1/2})$, we obtain
\begin{equation}
\label{eq:GinzburgParamSG}
G(b\,\xi) \propto\frac{\pRSB^{D/2-4}}{M} \int_0^{\infty} \frac{\mathrm{d}\alpha}{\alpha^{D/2-3}} \exp\left(-\frac{b^2}{4\alpha}-\alpha\right)\;.
\end{equation}
Note that the prefactor in Eq.~\eqref{eq:GinzburgParamSG} diverges in the critical region for $D<8$, leading to an upper critical dimension $\DU\geq 8$, in agreement with the upper critical dimension found in the BL expansion approaching the $T=0$ critical point from the PM phase in Ref.~\cite{angelini2022unexpected}.

\section{Conclusions}
\label{sec:Conclusions}

Concluding, we have analyzed the RSB phase of a SG model with external field on a BL at $T=0$. Despite the MF nature of the problem, the complete solution in the broken phase is not known and until now the differences with the much more understood RSB phase on the FC model were never clearly identified. In this paper, we highlight a crucial difference between the BL and the FC models: in the BL at $T=0$, thanks to local fluctuations due to finite connectivity, the RSB phase does not take place homogeneously on the whole system. In fact, there exists some spins that feel a unique local field in the many RSB LGS: in practice, they continue to behave as in the RS phase. This phenomenon has practical consequences on given instances of the problems: the frozen spins are much less involved in the avalanches that can be produced by a small perturbation of the GS and bring the system to another LGS. The largest avalanches involve the unfrozen spins with higher probability.

Our results are based on the computation of the extremes of the distribution of local fields on all the LGSs. The equations for the extremes do not depend on the number of RSB steps and on the corresponding RSB parameters. The fraction $p$ of open distributions that we compute is an upper bound to the actual number of unfrozen spins, as the proper reweighting of the LGSs could eventually close some open distributions, thus making the appearance of the RSB effects restricted to an even smaller fraction of spins.

Frozen and unfrozen spins can be identified also in the RFIM, with the latter being the only responsible for the spontaneous arising of a non-zero global magnetization. 

The role of the unfrozen spins in the RSB phase turns out to be crucial also in understanding how the MF behavior gets modified passing from the BL to a finite dimensional model.
Making use of the $M$-layer expansion, we have been able to identify the upper critical dimension $\DU=8$, that is in perfect agreement with what has been found approaching the transition from the PM phase \cite{angelini2022unexpected}.

We hope that the deeper understanding of the RSB phase obtained will be useful to design new optimization algorithms for the computation of the LGSs in polynomial time.
\begin{acknowledgements}
This research has been supported by the European Research Council under the European Union Horizon2020 research and innovation program (grant No.~694925 -- Lotglassy, G.~Parisi).
\end{acknowledgements}

\appendix

\section{The Cavity method at $T=0$}
\label{sec:kRSBansatz}
Below the de Almeida-Thoules (dAT) line, the phase space breaks into many states, and the correct solution of the Bethe lattice SG is supposed to require infinite steps of RSB \cite{Mezard2003,Franz2000,de2018computation}. We recall that the dAT line, that corresponds to the curve $\hat\sigma_H(T)$ in the $T-\sigma_H$ plane on which the spin glass susceptibility diverges, on the infinite tree is proved \cite{parisi2014diluted} to coincide with the locus of the points such that the following homogeneous linear integral equation with $k=2$ admits a non-zero solution $g(u)$:
\begin{multline}
\label{eq:CritPointCond}
g(u)=M\mathds{E}_{J,H}\int\dd hP_{M-1}(h)\,\dd u'\,g(u')\times\\\times \delta\left(u-\tilde{u}_{\beta}(J,H+u'+h)\right)\left(\frac{\dd\tilde{u}_{\beta}(H+u'+h)}{\dd H}\right)^k,
\end{multline}
where $\tilde{u}_{\beta}$ is the filter function at generic inverse temperature $\beta$ (see Ref.~\cite{Mezard2001}), $M=z-1$, and
\begin{multline}
P_{M-1}(h)=\mathds{E}_H\int\left[\prod_{i=1}^{M-1}Q_{RS}(u_i)\,\dd u_i\right]\times\\\times\delta\left(h-H-\sum_{i=1}^{M-1}u_i\right).
\end{multline}
By taking the zero temperature limit of \eqref{eq:CritPointCond} it is possible to compute the zero temperature critical point $\lim_{T\rightarrow 0}\hat{\sigma}_H(T):=\hat{\sigma}_H$. 
For $T=0$ Eq.~\eqref{eq:CritPointCond} becomes 
\begin{multline}
\label{eq:CritPointCondZeroT}
g(u)=M\mathds{E}_{J,H}\int\dd hP_{M-1}(h)\,\dd u'\,g(u')\times\\\times \delta\left(u-\textnormal{sgn}(J)(H+u'+h)\right)\mathds{1}(|H+u'+h|<|J|).
\end{multline}
Note that the dependence on $k$ disappeared, since the derivative on the RHS of \eqref{eq:CritPointCond} becomes the step function $\mathds{1}(|h|<|J|)$. 

Let us discuss the RSB solution on the Bethe lattice. We recall that in the SK model the breaking function $q=q(x,T,\sigma_H)$ is continuous and non-decreasing. In particular there are two points $x_m(T,\sigma_H)$ and $x_M(T,\sigma_H)$, such that for $x<x_m(T,\sigma_H)$:
\begin{equation}
q(x,T,\sigma_H)=q_m(T,\sigma_H),
\end{equation}
for $x>x_M(T,\sigma_H)$:
\begin{equation}
q(x,T,\sigma_H)=q_M(T,\sigma_H),
\end{equation}
and for $x_m(T,\sigma_H)<x<x_M(T,\sigma_H)$:
\begin{equation}
q_m(T,\sigma_H)<q(x,T,\sigma_H)<q_M(T,\sigma_H),\quad \dd q/\dd x>0,
\end{equation}
where $x_m$ and $x_M$, with $0\leq x_m\leq x_M\leq 1$, are called the minimum and maximum breaking parameters. In Refs.~\cite{vannimenus1981study,parisi1980simple}
it is shown that for small $T$ the inverse $x(q,\sigma_H,T)$ weakly depends on $\sigma_H$, and it can be expanded as follows:
\begin{equation}
x(q,T)= y(q)\,T+O\left(T^2\right),    
\end{equation}
where $y(q,T)$ is a regular function for $q<1$, and for $q$ close to one it is given by:
\begin{equation}
y(q)\approx (1-q)^{-1/2}.
\end{equation}
For a generic continuous transition from a RS to a fullRSB phase the value assumed by $x_M$ on the transition line is related to the ratio between two static six-point susceptibilities \cite{parisi2013critical}. By exploiting this property, in Ref.~\cite{parisi2014diluted} it is proven that for the Ising spin glass on the Bethe lattice the value of the breaking point at the zero temperature transition is  $x_M=1/2$. This result does not depend on the connectivity, and therefore it is also true for the SK model in the limit of $T\rightarrow 0$. 

In the case of the Bethe lattice SG we assume that, analogously to the the SK model, the RSB is characterized by a function $x(q,T,\sigma_H)$ such that for small $T$:
\begin{equation}
\label{eq:xDiQeT}
x(q,T,\sigma_H)=y(q,\sigma_H)\,T+O\left(T^2\right),
\end{equation}
where $y(q,\sigma_H)$ may be singular at $q=1$ (see Ref.~\cite{franz2000non}).  

In the presence of replica-symmetry-breaking the RS recursion relation \eqref{eq:RScavEQ} admits many solutions, each of them associated with a LGS of the system. Let us index the LGSs by $\alpha$, and denote by $\{h_{i\rightarrow j}\}^{\alpha}$ and $E^{\alpha}=E(\{h_{i\rightarrow j}\}^{\alpha})$, respectively, the $\alpha$-th solution and its Bethe energy, which is given by \cite{Mezard2003}:
\begin{equation}
\label{eq:EnergiaCavit}
E^{\alpha}= \sum_{i\in \mathcal V} \epsilon_i^{\alpha}-\sum_{(i,j)\in \mathcal E}\epsilon_{ij}^{\alpha},  
\end{equation}
where the $\epsilon_i$'s are the node energies:
\begin{equation}
\label{eq:CostoNodo}
\epsilon_i^{\alpha}=-\sum_{j\in\partial i}a_{J_{ji}}(h^{\alpha}_{j\rightarrow i})-\bigg|\sum_{j\in\partial i}\hat{u}_{J_{ij}}(h^{\alpha}_{j\rightarrow i})+H_i\bigg|,
\end{equation}
and the $\epsilon_{ij}$'s are the edge energies:
\begin{equation}
\label{eq:CostoEdge}
\epsilon_{ij}^{\alpha}=-\max_{\sigma_i,\sigma_j}\left[h^{\alpha}_{i\rightarrow j}\sigma_i+h^{\alpha}_{j\rightarrow i}\sigma_j+J_{ij}\sigma_i\sigma_j\right].
\end{equation}
In Eq.~\eqref{eq:CostoNodo} we introduced the filter function $a_J(h)$:
\begin{equation}
a_J(h)=\max\{|J|,|h|\}.
\end{equation}
At the 1RSB level the function $q(x,T,\sigma_H)$ is a step function that for a given $(T,\sigma_H)$ is defined by three numbers, namely two overlaps $q_1(T,\sigma_H)\geq q_0(T,\sigma_H)$, and the breaking parameter $0\leq x_1(T,\sigma_H)\leq 1$:
\begin{equation}
\begin{split}
q(x,T,\sigma_H)&=q_0(T,\sigma_H),\quad \textnormal{if} \quad x\leq x_1(T,\sigma_H)\\
q(x,T,\sigma_H)&=q_1(T,\sigma_H),\quad \textnormal{if} \quad x>x_1(T,\sigma_H).
\end{split}
\end{equation}
Under the assumption \eqref{eq:xDiQeT}, that in this case reads
\begin{equation}
x_1(T,\sigma_H)=y_1(\sigma_H)T+O(T^2),
\end{equation}
one has that the zero temperature limit of the Boltzman distribution $\mu(\underline{\sigma})$ can be written as follows: 
\begin{equation}
\mu(\underline{\sigma})=\frac{1}{Z}\sum_{\alpha}e^{-y_1E^{\alpha}}\mu_{\alpha}(\underline{\sigma}),
\end{equation}
where $\mu_{\alpha}$ is the measure associated with the LGS $\alpha$ \cite{mezard1987spin,Mezard2001,Mezard2003,mezard2009information}. Let us define for each directed edge $i\rightarrow j$ the distribution (population) $P_{i\rightarrow j}^{(1)}(h_{i\rightarrow j})$ of the cavity field $h_{i\rightarrow j}$: 
\begin{equation}
P^{(1)}_{i\rightarrow j}(h_{i\rightarrow j})=\frac{1}{Z}\sum_{\alpha}e^{-y_1E^{\alpha}}\delta(h_{i\rightarrow j}-h_{i\rightarrow j}^{\alpha}).
\end{equation}
In general $P^{(1)}_{i\rightarrow j}$ fluctuates from edge to edge. In the large graph limit the distribution of these populations (population of populations) defines the local order parameter $\mathcal{P}_{(1)}(P_{i\rightarrow j}^{(1)})$. Within the 1RSB cavity method it is possible to write the following distributional equation for $\mathcal{P}_{(1)}$ \cite{Mezard2003,mezard2009information}:
\begin{widetext}
\begin{equation}
\label{eq:recursionDistrFields}
P^{(1)}\stackrel{d}{=}\frac{1}{z_1[\{P^{(1)}_i\};y_1]}\int\left[\prod_{i=1}^{M} P^{(1)}_i\dd h_i\right]\,\exp{\left(y_1\sum_{i=1}^Ma_{J_i}(h_i)\right)}\delta\left(h-H-\sum_{i=1}^{M}\hat{u}_{J_i}(h_i)\right),
\end{equation}
\begin{equation}
\label{eq:zShift}
z_1[\{P^{(1)}_i\};y_1]=\int\left[\prod_{i=1}^{M} P^{(1)}_i\dd h_i\right]\,\exp{\left(y_1\sum_{i=1}^Ma_{J_i}(h_i)\right)}:=\exp\bigg(-y_1\Delta F_{iter}^{(1)}(\{P^{(1)}_i\};y_1)\bigg),
\end{equation}
\end{widetext}
where $\{\bullet_i\}:=\{\bullet_1,\dots,\bullet_M\}$, and the sign of equality in distribution ``$\stackrel{d}{=}$'' means that:
\begin{equation}
H\sim P_H,\,\, J_1,\ldots,J_M\,\overset{\text{i.i.d.}}{\sim}\,P_J,\,\, P_1^{(1)},\ldots,P_{M}^{(1)}\,\overset{\text{i.i.d.}}{\sim}\,\mathcal{P}_{(1)}.
\end{equation}
Equation \eqref{eq:recursionDistrFields} is usually called 1RSB energetic cavity equation. By changing the breaking parameter it is possible to study the statistical properties of the cavity field conditioning to a precise energy scale. In order to obtain the cavity fields at the scale of the dominant LGS one has to choose $y_1$ in such a way as to maximise the 1RSB free energy functional \cite{mezard1987spin,Mezard2003}:
\begin{equation}
\label{eq:1RSBfree}
\phi^{(1)}(\mathcal{P}_{(1)},y_1)=\mathbb E\left\{\Delta F^{(1)}_n-\frac{z}{2}\Delta F^{(1)}_e\right\},
\end{equation}
where $\mathbb E$ is the average w.r.t.\ $\mathcal{P}_{(1)},P_H$ and $P_J$.
The free energy of the node $\Delta F_n^{(1)}$ and of the edge $\Delta F_e^{(1)}$ entering \eqref{eq:1RSBfree} are respectively defined by:
\begin{widetext}
\begin{equation}
\label{eq:nodeFreeEn}
\exp\left(-y_1\Delta F_n^{(1)}(P_1^{(1)},\ldots,P_{z}^{(1)};y_1)\right)=\int\left[\prod_{i=1}^{z}P_i^{(1)}\dd h_i\right]\exp{\left(-y_{1}\epsilon_n(h_1,\dots,h_z;J,H)\right)},
\end{equation}
\begin{equation}
\label{eq:edgeFreeEn}
\exp\left(-y_1\Delta F_e^{(1)}(P_1^{(1)},P_2^{(1)};y_1)\right)=\int\left[\prod_{i=1}^{2}P_i^{(1)}\dd h_i\right]\exp{\left(-y_{1}\epsilon_e(h_1,h_2;J)\right)}.
\end{equation}
\end{widetext}
The breaking procedure can be formally iterated at any finite step $k$ of RSB (see Refs.~\cite{mezard2009information,semerjian2008freezing}). Let us call $R_0$ the set of probability distributions over the space of cavity fields $h$, and $R_{k+1}$ the space of probability distributions over $R_k$. In the following we call a distribution belonging to $R_r$ an $r$-distribution (or $r$-population). At a generic finite step $k$ the order parameter $\mathcal{P}_{(k)}$ becomes a $k$-distribution satisfying the following recursion equation:
\begin{equation}
\label{eq:fullRSBDistributional}
P^{(k)}(P^{(k-1)})\stackrel{d}{=}\mathcal{O}^{(k)}[P^{(k-1)};\{P_i^{(k)}\},y_1,\dots,y_k],
\end{equation}
where the sign of equality in distribution means that:
\begin{equation}
H\sim P_H,\,\, J_1,\ldots,J_M\,\,\overset{\text{i.i.d.}}{\sim}\,P_J,\,\, P_1^{(k)},\ldots,P_{M}^{(k)}\,\overset{\text{i.i.d.}}{\sim}\,\mathcal{P}_{(k)}
\end{equation}
and the function $\mathcal{O}^{(k)}$, taking as arguments $M$ $k$-distributions
\begin{equation}
\{P^{(k)}_i\}:=\{P^{(k)}_1,\dots,P^{(k)}_M\},   
\end{equation}
evaluates a new $k$-distribution $P^{(k)}$.

The definition of $\mathcal{O}^{(k)}$ is iterative, i.e.\ the function at level $k$ is defined in terms of the function at level $k-1$, and depends on the value of the $k$ breaking parameters:
\begin{equation}
0\leq y_1 \leq y_2\leq \dots\leq y_k\leq \infty,
\end{equation}
that, analogously to the 1RSB case, are obtained from the finite temperature breaking parameters $x_{r}$'s:
\begin{equation}
0\leq x_1\leq\dots\leq x_k\leq 1,
\end{equation}
by taking the following limit
\begin{equation}
y_{r}(\sigma_H)=\lim_{T\rightarrow 0}x_r(T,\sigma_H)/T,\quad r\in\{1,\dots,k\}.
\end{equation}
The iterative definition for $\mathcal{O}^{(k)}$ is the following:
\begin{widetext}
\begin{multline}
\label{eq:generalkRSBfunction}
\mathcal{O}^{(k)}[P^{(k-1)};\{P_i^{(\ell)}\},y_1\dots,y_k]=
\\=\frac{1}{z_{k}[\{P_i^{(k)}\},y_1,\dots,y_k]}\int\left[\prod_{i=1}^{M}P_i^{(k)}\dd P^{(k-1)}_i\right]\left(z_{k-1}[\{P_i^{(k-1)}\},y_2,\dots,y_k]\right)^{y_1/y_2}\times\\\times \delta\left[P^{(k-1)}(P^{(k-2)})-\mathcal{O}^{(k)}[P^{(k-2)};\{P_i^{(k-1)}\},y_2\dots,y_k]\right],
\end{multline}
\begin{multline}
\label{eq:EnergyShift}
z_{k}[\{P_i^{(k)}\},y_1,\dots,y_k]:=\exp\bigg(-y_1\Delta F^{(k)}_{iter}(\{P_i^{(k)}\},y_1,\dots,y_k)\bigg)=\\
=\int\left[\prod_{i=1}^{M}P_i^{(k)}\dd P^{(k-1)}_i\right]\left(z_{k-1}[\{P_i^{(k-1)}\},y_2,\dots,y_k]\right)^{y_1/y_2}=\\=\int\left[\prod_{i=1}^{M}P_i^{(k)}\dd P^{(k-1)}_i\right]\exp\bigg(-y_1\Delta F^{(k-1)}_{iter}(\{P_i^{(k-1)}\},y_2,\dots,y_k)\bigg).
\end{multline}
\end{widetext}
We want to observe that the $z_{k-1}$'s, the so called reweighting factors, are strictly positive, indeed we can write:
\begin{equation}
\label{eq:lowerBoundRewFact}
\left(z_{k-1}[\{P_i^{(k-1)}\},y_2,\dots,y_k]\right)^{y_1/y_2}\geq e^{y_1}.
\end{equation}
Analogously to the $k=1$ case, it is possible to define a $k$-RSB free energy functional $\phi^{(k)}$ depending on $\mathcal{P}_{(k)}$ and $y_1,\dots,y_k$, such that the cavity Eq.~\eqref{eq:fullRSBDistributional} is equivalent to the stationarity condition of $\phi^{(k)}$ w.r.t.\ to infinitesimal variations of $\mathcal{P}_{(k)}$. The $k$-RSB free energy functional is defined as follows:
\begin{equation}
\phi^{(k)}(\mathcal{P}_{(k)},y_1,\dots,y_k)=\mathbb E\left\{\Delta F_{n}^{(k)}-\frac{z}{2}\Delta F_{e}^{(k)}\right\},
\end{equation}
where $\mathbb E$ is the average w.r.t.\ $\mathcal{P}_{(k)}$, $P_H$ and $P_J$. The node term $\Delta F_{n}^{(k)}$ and the edge term $\Delta F^{(k)}_{e}$ are respectively given by the following recursion:
\begin{widetext}
\begin{multline}
\label{eq:Fn1}
\exp\left(-y_{1}\Delta F_{n}^{(k)}(P_1^{(k)},\ldots,P_{z}^{(k)};y_1,\dots,y_k)\right)=\\=\int\left[\prod_{i=1}^{z}P_i^{(k)}\dd P_i^{(k-1)}\right]\exp{\left(-y_{1}\Delta F_{n}^{(k-1)}(P_1^{(k-1)},\ldots,P_{z}^{(k-1)};y_2,\dots,y_k)\right)},
\end{multline}
\begin{multline}
\label{eq:Fn2}
\exp\left(-y_1\Delta F^{(k)}_{e}(P_1^{(k)},P_2^{(k)};y_1,\dots,y_k)\right)=\\=\int\left[\prod_{i=1}^{2}P_i^{(k)}\dd P_i^{(k-1)}\right]\exp{\left(-y_1\Delta F^{(k-1)}_{e}(P_1^{(k-1)},P_2^{(k-1)};y_2,\dots,y_k)\right)}.
\end{multline}
\end{widetext}
This construction can be applied to generic models defined on sparse graphs \cite{mezard2009information}. Usually the choice of $k$ and of the breaking parameters is made heuristically. In particular for a given $k$ the breaking parameters $y_1\leq\dots\leq y_k$ are chosen in such a way as to maximise $\phi^{(k)}$. In some cases indeed it has been proven for a generic temperature that the RSB free energy is a lower bound of the actual free energy of the system \cite{panchenko2004bounds,franz2003replica,guerra2003broken}, and it is conjectured that this property should hold in general \cite{mezard2009information}. Since the $k$-RSB ansatz incorporates as special cases all possible smaller level of RSB, the solution of the fullRSB ansatz ($k\uparrow\infty$), after the extremization over the breaking parameters, should provide the right value of $k$. This approach however is practically unfeasible since in most of the cases the numerical solution of the RSB equations becomes complicated already for $k=2$. Usually a plausibility check of the validity of the RS and the 1RSB descriptions is performed via a local stability analysis \cite{montanari2004instability}.

\section{The extreme values of the cavity field}
\label{sec:DistrExtrApp}
In this appendix we derive the recursive equation for the distribution of the maximum and minimum values taken by the local cavity field.
\begin{figure}
\centering
\includegraphics[width=0.8\columnwidth]{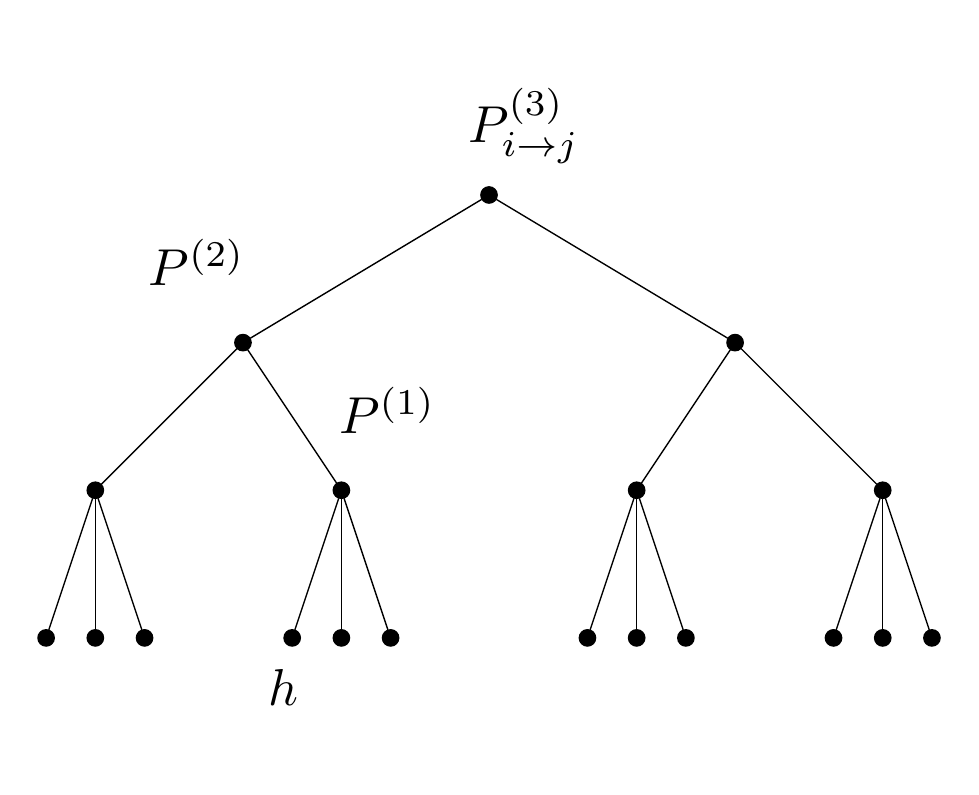}
\caption{
Representation of an RSB tree for $k=3$. Each node above the leaves is associated with a population. The leaves represent the value assumed by the cavity field on the directed edge $i\rightarrow j$ on the different states. We want to stress that the finite connectivity of the Bethe lattice implies a spacial dependence of the order parameter.\label{fig:kRSB}
} 
\end{figure}

At the $k$-th step of RSB the statistical properties of the cavity field on a given site can be represented by a tree. On this tree the leaves correspond to the LGSs, and are associated with a realisation of the cavity field. The generic node at distance $\ell$ from the nearest leaf corresponds to an $\ell$-population $P^{(\ell)}$ (see Fig.~\ref{fig:kRSB}). In the following we denote by $\mathcal{L}(P^{(\ell)})$ the set of leaves that are descendant of $P^{(\ell)}$, i.e.\ the leaves at distance $\ell$ from $P^{(\ell)}$. Let us define for $\ell=1,\ldots,k$ the extremes $\boldsymbol{\mathrm{h}}^{+}(P^{(\ell)})$ and $\boldsymbol{\mathrm{h}}^{-}(P^{(\ell)})$, conditioned to a population $P^{(\ell)}$:
\begin{equation}
\boldsymbol{\mathrm{h}}^{+}(P^{(\ell)})=\max\,\mathcal{L}(P^{(\ell)})\,,\quad \boldsymbol{\mathrm{h}}^{-}(P^{(\ell)})=\min\,\mathcal{L}(P^{(\ell)}).
\end{equation}
Equivalently $\boldsymbol{\mathrm{h}}^{+}(P^{(\ell)})$ and $\boldsymbol{\mathrm{h}}^{-}(P^{(\ell)})$ represent the maximum and minimum values of the cavity fields conditioned to $P^{(\ell)}$. 
On the directed edge $i\rightarrow j$ the maximum $h^+_{i\rightarrow j}$ and minimum $h^-_{i\rightarrow j}$ values of the extremes are given by:
\begin{equation}
h^+_{i\rightarrow j}=\max\,\mathcal{L}(P^{(k)}_{i\rightarrow j})\,,\quad h^-_{i\rightarrow j}=\min\,\mathcal{L}(P^{(k)}_{i\rightarrow j}).
\end{equation}
The couple $(h^+_{i\rightarrow j},h^-_{i\rightarrow j})$ depends on the directed edge, and its probability distribution $\mathcal{P}(h^+,h^-)$
is defined by the k-RSB order parameter $\mathcal{P}_{(k)}$ as follows:
\begin{multline}
\label{eq:ExtremesDef}
\mathcal{P}(h^{+},h^{-})=\int \mathcal{P}_{(k)}\dd P^{(k)}\times\\\times\delta\left[h^{+}-\boldsymbol{\mathrm{h}}^{+}(P^{(k)})\right]\,\delta\left[h^{-}-\boldsymbol{\mathrm{h}}^{-}(P^{(k)})\right].
\end{multline}
We want to show that it is possible to write a closed equation for $\mathcal{P}(h^{+},h^{-})$
that does not depend on the parameters defining the symmetry breaking, implying that $\mathcal{P}(h^{+},h^{-})$ does not depend on $k$ and on $y_1,\dots,y_k$. Indeed denoting by $\mathcal{S}(P^{(\ell)})$ the sons of $P^{(\ell)}$, i.e.\ the neighbouring nodes of $P^{(\ell)}$ whose nearest leaf is at distance $\ell-1$, we can write:
\begin{equation}
\label{eq:hMenohPiu}
\begin{split}
h^{+}&\stackrel{d}{=} H+\sum_{i=1}^{M}\max_{\mathcal{L}(P_i^{(k)})}\hat{u}_{J_i}(h_i),\\
h^{-}&\stackrel{d}{=} H+\sum_{i=1}^{M}\min_{\mathcal{L}(P_i^{(k)})}\hat{u}_{J_i}(h_i).  
\end{split}  
\end{equation}
Equations \eqref{eq:hMenohPiu} are obtained, in order, starting from the definition \eqref{eq:ExtremesDef}, using the RSB cavity equations \eqref{eq:fullRSBDistributional}, and then by applying iteratively the following property
\begin{equation}
\label{eq:defFunctExtrem}
\begin{split}
\boldsymbol{\mathrm{h}}^{+}(P^{(\ell)})&=
\max_{\mathcal{S}(P^{(\ell)})}
{\boldsymbol{\mathrm{h}}^{+}(P^{(\ell-1)})}\\ \boldsymbol{\mathrm{h}}^{-}(P^{(\ell)})&=
\min_{\mathcal{S}(P^{(\ell)})}
{\boldsymbol{\mathrm{h}}^{-}(P^{(\ell-1)})},
\end{split}
\end{equation}
that is a consequence of the positivity of the reweighting factors (see Eq.~\eqref{eq:lowerBoundRewFact}). At this point Eq.~\eqref{eq:hMenohPiu} can be closed observing that the minimum and the maximum of $\hat{u}_{J_i}(h_i)$ do not depend on the values of the fields taken on all the leaves, but only on $h_i^{+}$ and $h_i^{-}$, being $\hat{u}$ monotonic (see Eq.~\eqref{eq:filterFunctU}). In particular by defining the ordering functions
\begin{equation}
\label{eq:ordFunc}
\begin{split}
f^{(+)}_J(h^{+},h^{-})&=\max{\big\{\hat{u}_J(h^{+}),\hat{u}_J(h^{-})\big\}},\\ f^{(-)}_J(h^{+},h^{-})&=\min\big{\{\hat{u}_J(h^{+}),\hat{u}_J(h^{-})\big\}},
\end{split}
\end{equation}
we obtain equation:
\begin{multline}
\label{eq:recSupport}
\mathcal{P}(h^{+},h^{-})=\mathds{E}_{H}\int\left[\prod_{i=1}^{M}\mathcal{Q}(u^{+}_i,u^{-}_i)\,\dd u^{+}_i\dd u^{-}_i\right]\times\\\times\delta\left(h^{+}-H-\sum_{i=1}^{M}u^{+}_i\right)\,\delta\left(h^{-}-H-\sum_{i=1}^{M}u^{-}_i\right),
\end{multline}
where we introduced the joint distribution of the extremes $\mathcal{Q}(u^{+},u^{-})$ of the $u$ fields:
\begin{multline}
\label{eq:RecDistrUExtremes}
\mathcal{Q}(u^{+},u^{-})=\mathds{E}_{J}\int \dd h^{+}\dd h^{-}\mathcal{P}(h^{+},h^{-}) \times\\\times\delta\big(u^{+}-f^{(+)}_J(h^{+},h^{-})\big)\,\delta\big(u^{-}-f^{(-)}_J(h^{+},h^{-})\big).
\end{multline}
As already discussed in section \ref{sec:RSBT0}, the extremes do not depend on $k$ and on the Parisi breaking parameters $y_1,\ldots,y_k$. Note that when $\mathcal{P}(h^{+},h^{-})$ concentrates on the line $h^{+}=h^{-}$, meaning that the $k$-populations reduce to a single number, Eq.~\eqref{eq:recSupport} equals the replica symmetric equation \eqref{eq:RScavEQ}. 

All these arguments can repeated also for $T>0$. In this case however there is a crucial difference, namely the finite-temperature filter-function (see Ref.~\cite{Mezard2001}):
\begin{equation}
\tilde{u}_{\beta}(J,h)=\frac{1}{\beta}\arctanh(\tanh(\beta J)\tanh(\beta h))
\end{equation}
is injective, and then below $\hat\sigma_H(T)$ the probability $p$ of drawing an open couple is always strictly equal to one. 

The recursions \eqref{eq:recSupport} and \eqref{eq:RecDistrUExtremes}, which are closed equations for the joint probability distribution of the extremes, can be also rewritten in a single-instance version. Indeed consider a graph $G(\mathcal{V},\mathcal{E})$ equipped with a set of couplings $\{J_{e}\}_{e\in\mathcal{E}}$ and external fields $\{H_i\}_{i\in\mathcal{V}}$. Let us associate with each directed edge $i\rightarrow j$, $(i,j)\in\mathcal{E}$, a couple of $h$-extremes $(h^+_{i\rightarrow j},h^-_{i\rightarrow j})$, and a couple of $u$-extremes $(u^+_{i\rightarrow j},u^-_{i\rightarrow j})$. From \eqref{eq:recSupport} and \eqref{eq:RecDistrUExtremes} we write 
\begin{equation}
\label{eqSingleInstEqExtr1}
u^{+}_{i\rightarrow j}=f^{(+)}_J(h^{+}_{i\rightarrow j},h^{-}_{i\rightarrow j}),\quad u^{-}_{i\rightarrow j}=f^{(-)}_J(h^{+}_{i\rightarrow j},h^{-}_{i\rightarrow j}),
\end{equation}
where
\begin{equation}
\label{eqSingleInstEqExtr2}
h^+_{i\rightarrow j}=H_i+\sum_{k\in\partial i\setminus j}u^{+}_{k\rightarrow i},\quad 
h^-_{i\rightarrow j}=H_i+\sum_{k\in\partial i\setminus j}u^{-}_{k\rightarrow i}.
\end{equation}
Equations \eqref{eqSingleInstEqExtr1} and \eqref{eqSingleInstEqExtr2} can be solved by iteration with a message-passing algorithm. In this way, for a given realization of the disorder, it is possible to predict if a spin is closed or open. This turns to be particularly informative in relation to the phenomenon of the spin avalanches, as discussed in section \ref{sec:avalanches}. 

We want to conclude the appendix with a comment about the limit $T\rightarrow 0$. We introduced the extremes starting from the energetic RSB cavity equations. A complete control of the limit $T\rightarrow 0$ would require the study of small temperature perturbations, in order to check if the entropic contributions are important. Indeed the assumption \eqref{eq:xDiQeT} of regularity of the $T\rightarrow 0$ limit, corresponds to require that the energetic landscape at $T=0$ has the same form of the free energy landscape for $T>0$, or analogously that after the introduction of a small temperature, the entropic effects do not produce dramatic changes to the relative weights of the states (see Ref.~\cite{marinari2000replica}). Here we want to observe that since the entropic effects can only result in a reweighting of the LGS, they cannot open the closed populations. Therefore we expect that a detailed analysis of these contributions, that will be the object of a future investigation, at least should not affect the prediction of the existence of closed spins in the RSB phase.

\section{Critical behavior of the extremes}
\label{sec:ExpansionCloseToCrit}
In this appendix we derive the asymptotic behavior of the extremes close to the critical point. This can be done by exploiting the double effect of $p$ becoming small together with the width of the open populations.   

Let us denote by $\epsilon=\hat{\sigma}_H-\sigma_H$ the distance from the critical point. 
It is useful to represent the couple of fields in terms of the width $\Delta$, and the median value $h$:
\begin{equation}
h^+=h+\Delta/2,\quad h^-=h-\Delta/2,
\end{equation}
and to distinguish between open and closed populations:
\begin{equation}
Q(u,\Delta)=Q_o(u,\Delta)+Q_c(u)\,\delta(\Delta)
\end{equation}
\begin{equation}
P(h,\Delta_h)=P_o(h,\Delta_h)+P_c(h)\,\delta(\Delta_h).
\end{equation}
At variance with \eqref{eq:decompOrdPar}, here it is convenient to use the following normalizations:
\begin{equation}
\int Q_o(u,\Delta)\,\dd u\,\dd\Delta=p,\quad \int  P_o(h,\Delta_h)\,\dd h\,\dd\Delta_h=\hat{p},
\end{equation}
where $\hat{p}=1-(1-p)^{z-1}$ is the probability of drawing an $h$-couple $(h^+,h^-)$ with $\Delta_h>0$.  

Equation \eqref{eq:RecDistrUExtremes} can be rewritten in the following form:
\begin{equation}
\label{eq:QoQcNewNot1}
Q_o(u,\Delta) = \int F_1(u,\Delta,h,\Delta_h) P_o(h,\Delta_h)\, dh \, d\Delta_h
\end{equation}
\begin{multline}
\label{eq:QoQcNewNot2}
Q_c(u) = \int G_0(u,h) P_c(h)\, dh+\\ + \int F_2(u,h,\Delta_h)P_o(h,\Delta_h)\, dh\,d\Delta_h,  
\end{multline}
where we introduced $G_0$:
\begin{equation}
\label{eq:G0}
G_0(u,h)\equiv \mathds{E}_J\,\delta\big(u-\hat{u}_{J}(h)\big),
\end{equation}
and the two distributions $F_1,F_2$. $F_1$ is defined by
\beq
\label{eq:defF1}
F_1 \equiv \mathds{E}_J\,\delta\big(u-\text{sgn}(J)\, h\big)\,\delta(\Delta-\Delta_h) \,\mathds{1}\big(|h|<|J|\big) + \Delta F_1 \ ,
\eeq
where $\mathds{1}(A)$ is the indicator function of the set $A$. Note that we have $\Delta F_1 \neq 0$ if and only if
\begin{equation}
\big||J|-h\big|<\frac{\Delta_h}{2}\quad\text{or}\quad\big||J|+h\big|<\frac{\Delta_h}{2},
\end{equation}
and since $\Delta_h$ is typically small we consider $\Delta F_1$ as a perturbation. $F_2$ is given by
\beq
\label{eq:defF2}
F_2 \equiv \mathds{E}_J\,\delta\big(u-\text{sgn}(J)\,\text{sgn}(h)\big)\,\mathds{1}\left(|J|+\frac{\Delta_h}{2}<|h|\right),
\eeq
and this can also be written as
\beq
F_2 \equiv \mathds{E}_J\,\delta\big(u-\text{sgn}(J)\,\text{sgn}(h)\big)\mathds{1}\big(|h|>|J|\big)+\Delta F_2, 
\eeq
where again $\Delta F_2$ can be treated as a perturbation. Note that at the leading order, independently of $\Delta$, either $|h|<|J|$ and the open population remains open with the same $\Delta$, or $|h|>|J|$ and the population closes. 

In the following if not necessary we omit the dependence of the distributions on $u,\Delta,h,\Delta_h$, and we use the following simplified notation:
\begin{align}
\label{eq:Qoexact}
Q_o & = \mathbf{F}_1 P_o \\
\label{eq:Qcexact}
Q_c & = \mathbf{G}_0 P_c + \mathbf{F}_2 P_o \\
\label{eq:Pcexact}
P_c & = (Q_c)^M P_H \\
\label{eq:Poexact}
P_o & = (Q_a+Q_c)^MP_H -(Q_c)^MP_H,
\end{align}
in which we denoted the convolution between distributions as a product, and we introduced the linear integral operators $\mathbf{G}_0$,$\mathbf{F}_1$,$\mathbf{F}_2$ associated with \eqref{eq:G0}, \eqref{eq:defF1} and \eqref{eq:defF2}, namely:
\begin{align}
\mathbf{G}_0 P_c&\equiv\int G_0(u,h) P_c(h)\, dh,\\ 
\mathbf{F}_1P_o&\equiv\int F_1(u,\Delta,h,\Delta_h)P_o(h,\Delta_h)\, dh \, d\Delta_h,\\
\mathbf{F}_2P_o&\equiv \int F_2(u,h,\Delta_h)P_o(h,\Delta_h).
\end{align}
As an example of the simplified notation introduced in equations \eqref{eq:Qoexact}-\eqref{eq:Poexact} consider 
\begin{multline}
Q_o^2Q_c^{M-2}P_H\equiv(Q_o^2Q_c^{M-2}P_H)(h,\Delta_h)=\\=\int\left[\prod_{i=1}^{M-2}Q_c(u_i)\dd h_i\right]\left[\prod_{i=1}^2\,Q_o(u_i',\Delta)\,\dd u_i'\dd \Delta_i\right]\times\\\times\delta\left(h-H-\sum_{i=1}^{M-2}u_i-u_1'-u_2'\right)\delta(\Delta_h-\Delta_1-\Delta_2).
\end{multline}
Before going on, it is worth introducing the Laplace transform of $Q_o$ with respect to $\Delta$:
\beq
q_o(s) \equiv \int e^{s \Delta} Q_o(u,\Delta)\, d \Delta=(\mathbf{L}Q_o)(s).
\eeq
In the following we make explicit only the dependence on $s$, and the product of Laplace transforms has to be intended as convolutions on all variables but $s$. 

For the computation of $q_o(s)$ we apply $\mathbf{L}$ to \eqref{eq:Qoexact}:
\begin{equation}
\label{la13}
q_o(s)=\mathbf{F}_0\, p_o(s)+(\mathbf{L}\mathbf{\Delta F}_1 P_o)(s),
\end{equation}
where 
\begin{equation}
F_0\equiv\mathds{E}_J\,\delta\big(u-\text{sgn}(J)\, h\big)\,\mathds{1}\big(|h|<|J|\big).
\end{equation}
Observe how the leading contribution to $\mathbf{F}_1$ commutes with $\mathbf{L}$ since it acts as the identity on $\Delta$. From the equation \eqref{eq:Poexact} for $P_o$ we have:
\begin{equation}
\label{la14}
\begin{split}
p_o(s) = & M q_o(s)\,Q_c^{M-1}P_H+\\ &+\frac{M(M-1)}{2} q_o(s)^2\,Q_c^{M-2}P_H+\dots
\end{split}
\end{equation}
Note that the convolution has been transformed into products. By writing:
\begin{align}
\label{eq:expancloscrit}
P_H & = P_H^{crit}+\delta P_H
\\
Q_c & = Q^{crit}+\delta Q_c
\end{align}
we can expand \eqref{la14} close to the critical point:
\begin{align}
\label{eq:po}
p_o(s) = & M q_o(s) (Q^{crit})^{M-1}P_H^{crit}+
\\
& + M q_o(s)\,(Q^{crit})^{M-1}\delta P_H+
\label{eq:neglectermdeltaPH}
\\
& + M (M-1) \, q_o(s)\,\delta Q_c (Q^{crit})^{M-2}P_H^{crit}+
\\
& + \frac{M (M-1)}{2} \, q_o(s)^2\, (Q^{crit})^{M-2}P_H^{crit}+ \dots
\end{align}
At this point let us substitute \eqref{eq:po} into \eqref{la13}. We find that the leading term is given by
\begin{equation}
\label{eq:SymOp}
q_o(s)=M\,\mathbf{F}_0 (Q^{crit})^{M-1}P_H^{crit}q_o(s)+\dots,
\end{equation}
where the linear operator $M\,\mathbf{F}_0 (Q^{crit})^{M-1}P_H^{crit}$ sends a function $f(u)$ defined for $|u|<1$ into another function on the same space. Note that generically we work with functions that can have delta peak at $u=\pm 1$, but the operator naturally is always applied to functions that have no peaks by construction, indeed the variable $u$ in $Q(u,\Delta)$ cannot have peaks in $u=\pm 1$.
By definition the above operator has a (critical) eigenvalue that goes to one at the transition. We call $g^s_{crit}(u)$ and $g^d_{crit}(u)$ the corresponding left and right critical eigenvectors. 

Let us define
\beq
\label{eq:g(s)}
\int du \, g^s_{crit}(u)\,q_o(u,s) = g^s_{crit}\cdot q_o(s)\equiv g(s),
\eeq
and choose the normalizations:
\beq
g^s_{crit}\cdot g^d_{crit}=1,\quad \int du \, g^d_{crit}(u) = 1.
\eeq
In Eq.~\eqref{eq:g(s)} we introduced the symbol $\cdot$ to denote the integration on the variable $u$. Note that in the case of symmetric disorder $M\,\mathbf{F}_0 (Q^{crit})^{M-1}P_H^{crit}$ is a symmetric operator, and then: 
\beq
\label{eq:symDisgSgD}
g^s_{crit}(u)=\frac{g^d_{crit}(u)}{\int\dd u\, g^d_{crit}(u)^2}.
\eeq
Now by the definition \eqref{eq:g(s)} we write
\begin{equation}
\label{lead}
q_o(s)=g(s)\,g^d_{crit}+\dots,
\end{equation}
where the dots are the contributions of the subleading eigenvectors. In order to obtain an equation for $g(s)$, let us project \eqref{la13} on the critical left eigenvector $g^s_{crit}$:
\begin{equation}
\label{la13proj}
g(s)=g^s_{crit}\cdot\mathbf{F}_0\, p_o(s)+g^s_{crit}\cdot(\mathbf{L}\mathbf{\Delta F}_1 P_o)(s).
\end{equation}
On the LHS of Eq.~\eqref{la13proj} we simply have $g(s)$ from the definition \eqref{lead}. On the RHS the first contribution is obtained by expanding close to criticality:
\begin{equation}
\begin{split}
\mathbf{F}_0 p_o(s) \approx & g(s)+\frac{M(M-1)}{2}\,\mathbf{F}_0 q_o(s)^2 (Q^{crit})^{M-2}P_H^{crit}+\\&+ M (M-1)\, \mathbf{F}_0 q_o(s) (Q^{crit})^{M-2}P_H^{crit}\delta Q_c.
\end{split}
\end{equation}
Note that from Eq.~\eqref{lead} one has
\begin{equation}
\label{eq:margDistrMedianValField}
\underline{Q_a}(u)=g(0)\,g_{crit}^d+\dots,
\end{equation}
where the underline denotes the marginalisation over $\Delta$. As it is shown at the end of the appendix (see equations \eqref{eq:TotMargQP}-\eqref{eq:varQcexpla}), Eq.~\eqref{eq:margDistrMedianValField} allows us to write the variation of the distribution of the closed population:
\beq
\label{eq:variationClosedPop}
\delta Q_c=-g(0)\, g_{crit}^d+\dots
\eeq
We can thus write:
\beq
\label{eq:primoPezzo}
g^s_{crit}\cdot \mathbf{F}_0 p_o(s) = g(s)+B\, \left(g^2(s)-2g(s)g(0)\right)+\dots
\eeq
with
\beq
\label{eq:DefB}
B \equiv \frac{M(M-1)}{2}\, g^s \cdot {\bf F_0}  (g^d_{crit})^2 (Q^{crit})^{M-2}P_H^{crit}.
\eeq
In order to study the second term on the RHS of \eqref{la13proj},
\beq
g^s_{crit} \cdot ( {\bf L} \, {\bf \Delta F_1} P_o)(s),
\eeq
we write at leading order 
\beq
P_o= M Q_o (Q^{crit})^{M-1}P_H^{crit} +\dots
\eeq
in which by using Eq.~(\ref{lead}) one has
\beq
Q_o(u,\Delta)= g^d_{crit}(u)P(\Delta),
\eeq
and $P(\Delta)$ is the probability of $\Delta$. Note that 
\beq
\int P(\Delta)\,d\Delta=g(0)=p,
\eeq
and it is small close to the transition. We have
\beq
P_o(h,\Delta_h)= M g_1(h) \, P(\Delta_h) +\dots,
\label{Poleading}
\eeq
where 
\beq
\label{eq:defg1}
g_1(h)=(g^d_{crit} (Q^{crit})^{M-1}P_H^{crit})(h)
\eeq
is normalized to one, and can be evaluated in population dynamics.
Now we rewrite the operator $\Delta F_1$ as:
\beq
\label{eq:opDFterms}
\Delta F_1=\Delta F_1'+\Delta F_1''+\Delta F_1'''.
\eeq
The first term in \eqref{eq:opDFterms} takes into account those couples that were wrongly considered unaltered at leading order:
\begin{multline}
\label{eq:dFP}
\Delta F_1'=-\mathds{E}_J\,\bigg[\mathds{1}\left(0<|J|-|h|<\frac{\Delta_h}{2}\right)\times\\\times\delta\big(u-\text{sgn} (J)\,h\big)\,\delta(\Delta-\Delta_h)\bigg].
\end{multline}
The second term, taking into account couples that were wrongly considered to be closed at leading order, can be written as follows:
\begin{multline}
\label{eq:dFS}
\Delta F_1''=\mathds{E}_J\,\bigg[\mathds{1}\left(\big||J|-|h|\big|<\frac{\Delta_h}{2}\right)\times\\\times\delta\big(u-u_F\left(J,h,\Delta_h\right)\big)\,\delta\big(\Delta-\Delta_F\left(J,h,\Delta_h\right)\big)\bigg],
\end{multline}
where we defined
\begin{equation}
u_F\left(J,h,\Delta_h\right)=\frac{\text{sgn}(J\,h)}{2}\left(|J|+|h|-\frac{\Delta_h}{2}\right),
\end{equation}
and
\begin{equation}
\Delta_F\left(J,h,\Delta_h\right)=|J|-|h|+\frac{\Delta_h}{2}.
\end{equation}
The last term, $\Delta F_1'''$, takes into account the cases in which $h^+>|J|$ and $h^-<-|J|$, and therefore it is not relevant for the limit of small widths. 

Using definition (\ref{eq:dFP}) and (\ref{eq:dFS}) in the case of $|J|=1$, and writing at the leading order $P_o(h,\Delta_h)$
according to Eq.~(\ref{Poleading}) we find that:
\begin{multline}
\label{eq:leadingD}
g^s_{crit}\cdot(\mathbf{L}\mathbf{\Delta F_1} P_o)(s)\approx\\\approx 2\,B_2 \int  e^{s \Delta}\left(\int_{\Delta}^{\infty} P(\Delta')d \Delta'
 -\frac{\Delta}{2}P(\Delta )\right)d\Delta,
\end{multline}
with 
\beq
\label{eq:DefB2}
B_2= g^s_{crit}(1)g_1(1) M \, .
\eeq
Note that from the definition of \eqref{eq:defg1}:
\beq 
M\,g_1(v)=g^d(v),\quad |v|\leq 1,
\eeq 
therefore in the case of symmetric disorder, by using \eqref{eq:symDisgSgD}, we can write:
\beq
\label{eq:DefB2Sym}
B_2=\frac{g^d_{crit}(1)^2}{\int\dd u\, g^d_{crit}(u)^2}.
\eeq
From Eq.~\eqref{eq:leadingD}, expanding for small $\Delta$, we have:
\beq
\label{eq:ScalinoRSB}
g^s_{crit}\cdot(\mathbf{L}\mathbf{\Delta F_1} P_o)(s)= B_2 \left(2\,\frac{g(s)-g(0)}{s} -\dot{g}(s) \right).
\eeq
At this point by putting together the two contributions \eqref{eq:primoPezzo} and \eqref{eq:ScalinoRSB}
we obtain the equation:
\beq
\label{eq:EquazioneLaplace}
2 \frac{y(z)-1}{z} -\dot{y}(z)  +y^2(z)  -2 y(z) =0,
\eeq
where we changed variables as follows
\beq
z=\frac{s B g(0)}{B_2} \, , \ \  y(z)=\frac{g(s)}{g(0)}.
\eeq
Equation \eqref{eq:EquazioneLaplace} has the solution
\beq
\label{eq:lapTransfExp}
y(z)=\frac{1}{1-z},
\eeq
whose inverse Laplace transform is the exponential:
\beq
\label{eq:ExpLaw}
P(\Delta)=p \, \frac{1}{\Delta_{typ} }e^{-\Delta/\Delta_{typ}},
\eeq
with
\beq
\Delta_{typ}=\frac{B}{B_2}p.
\eeq
Equation \eqref{eq:lapTransfExp} is not the unique solution of \eqref{eq:EquazioneLaplace}, however other solutions have many poles. Indeed consider the algebraic equation obtained by computing the derivative of \eqref{eq:EquazioneLaplace} in zero. The non-uniqueness follows from the fact that the terms with $\ddot{y}(0)$ cancel out, leaving the second derivative unfixed. Interestingly this cancellation implies also that the algebraic equation cannot be satisfied if in \eqref{eq:EquazioneLaplace} there was a different linear term in $y(z)$, i.e.\ the solution would not exist. This means that in the derivation of \eqref{eq:EquazioneLaplace} it is consistent to consider $\epsilon=o(p)$, i.e.\ to neglect terms like \eqref{eq:neglectermdeltaPH}. 

At this point let us come back to \eqref{eq:variationClosedPop} and show how to obtain the variation $\delta Q_c$. Consider the total marginal probability distributions $Q$ and $P$ of, respectively, $u$ and $h$
\begin{equation}
\label{eq:TotMargQP}
\begin{split}
Q&=\underline{Q_o}+Q_c\\
P&=\underline{P_o}+P_c, 
\end{split}
\end{equation}
where, as before, the underline denotes the marginalisation over $\Delta$. Integrating \eqref{eq:Qoexact}-\eqref{eq:Poexact} with respect to $\Delta$, we find the following self-consistent equations for $Q$ and $P$:
\begin{align}
\label{eq:qoUnderqc1}
Q & = \mathbf{G}_0\,P+\underline{\mathbf{\Delta F}_1 P_o}+\mathbf{\Delta F}_2 P_o\\
\label{eq:qoUnderqc2}
P & = Q^M P_H,
\end{align}
that we want to study close to the critical point. Note that in the limit $\epsilon\downarrow 0$ we obtain the RS equations:
\begin{align}
\label{eq:RScavityeq}
Q & = \mathbf{G}_0\,P \\
P & = Q^M P_H^{\text{crit}}.
\end{align}
Now by writing: 
\begin{equation}
Q=Q^{\text{crit}}+\delta Q,\quad P=P^{\text{crit}}+\delta P,
\end{equation}
we can expand \eqref{eq:qoUnderqc1} and \eqref{eq:qoUnderqc2}
\begin{align}
\label{eq:qoqCvar1}
\delta Q & = \mathbf{G}_0\delta P+\underline{\mathbf{\Delta F}_1 P_o}+\mathbf{\Delta F}_2 P_o+\dots\\
\label{eq:qoqCvar2}
\delta P & = (Q^{crit})^{M}\,\delta P_H+M\delta Q\, (Q^{crit})^{M-1}P_H^{crit}+\dots
\end{align}
At the leading order the variations of $Q$ and $P$ are given by the sum of two contributions:  
\begin{equation}
\label{eq:qoqCvar3}
\delta Q=\delta Q_{RS}+\delta_p Q+\dots,
\end{equation}
where $\delta_p$ denotes the variation at fixed $\epsilon=0$, and $\delta Q_{RS}$ and $\delta P_{RS}$ are the variations of the RS distributions, i.e.\ the variations of $Q$ and $P$ at $p=0$. Since the RS equations \eqref{eq:RScavityeq} are regular at the transition, we should have
\begin{equation}
\label{eq:varQ0}
\delta Q_{RS}, \delta P_{RS}=O(\epsilon). 
\end{equation}
For the variations at $\epsilon=0$ we have
\begin{equation}
\label{eq:varEpsilon0}
\delta_p Q, \delta_p P=o(p),
\end{equation}
since $\underline{\mathbf{\Delta F}_1 P_o}+\mathbf{\Delta F}_2 P_o=o(p)$. At this point note that the variations of the marginals can be rewritten as follows
\begin{equation}
\delta Q=\underline{Q_o}+\delta Q_c,\quad \delta P=\underline{P_o}+\delta P_c,
\end{equation}
and that
\begin{equation}
\underline{Q_o},\delta Q_c=O(p).
\end{equation}
Putting all together, what we found is that the corrections to the total marginals $Q$ and $P$ are smaller than the corrections to the separate contributions coming from the open and closed distributions:
\begin{equation}
\label{eq:smallerQoUQc}
\underline{Q_o}+\delta Q_c=o(p),\quad  \underline{P_o}+\delta P_c=o(p), 
\end{equation}
implying, as anticipated in \eqref{eq:variationClosedPop}, that at the leading order 
\begin{equation}
\label{eq:varQcexpla}
\delta Q_c\approx -\underline{Q_o}.
\end{equation}

\section{Critical behavior of the RFIM}
\label{RFIMcritical}
In the last appendix the populations are expanded close to the critical point in terms of $p$. In order to complete the analysis we should find how $p$ scales with $\epsilon$, the distance from $\hat{\sigma}_H$. The strategy is to exploit the analogy with the RFIM, and in particular the fact that $n_{\text{unf}}=m$ (see Eq.~\eqref{eq:nUnfM}). For this purpose here we focus on the computation of $m$, the magnetization of the RFIM, as a function of $\epsilon$, and in appendix \ref{sec:CompQeps} on the computation of $n_{\text{unf}}$, the fraction of unfrozen spin in the SG, as a function of $p$.  

Let us use also for the RFIM the simplified notation introduced in appendix \ref{sec:ExpansionCloseToCrit}. For example we write the RS cavity equation,
\begin{multline}
Q(u)=\mathds{E}_H\int\left[\prod_{i=1}^{M}Q(u_i)\dd u_i\right]\times\\\times\delta\left(u-\hat{u}_1\left(H+\sum_{i=1}^Mu_i\right)\right),
\end{multline}
as follows 
\begin{equation}
\label{eq:RSsimplNotRFIM}
Q=\mathbf{R}_0\,Q^{M}P_H,   
\end{equation}
where 
\begin{equation}
(\mathbf{R}_0P)(u)=\int\dd h P(h)\,\delta\left(u-\hat{u}_{1}(h)\right).
\end{equation}
One expects that in the ferromagnetic phase a magnetized solution of the RFIM cavity equation develops continuously. This leads to an integral equation corresponding to \eqref{eq:CritPointCond} with $k=0$. Indeed let us rewrite the field distribution as the sum of a symmetric and an antisymmetric part:
\begin{equation}
Q(u)=S(u)+A(u).    
\end{equation}
There is always the solution $A(u)=0$, but close to the critical point, in order to study the magnetized solution, we can expand the cavity distribution as
\begin{equation}
\label{eq:DecompPRFIM}
Q(u)=Q_c(u)+\delta S(u)+ m \,f^d(u),
\end{equation}
where $Q_c(u)$ is $Q(u)$ at the critical point, $\delta S$ is the variation of $S$, and $f^d$ is the right antisymmetric eigenvector of the so called longitudinal operator $\mathbf{R}$ at the critical point: 
\begin{equation}
\label{eq:defGrassettoR}
\mathbf{R}=M\,\mathbf{R}_0\,Q_c^{M-1}P_H^{crit},  
\end{equation}
as can be seen by substituting \eqref{eq:DecompPRFIM} into \eqref{eq:RSsimplNotRFIM} and taking the limit $\epsilon\rightarrow 0$. Equivalently $f^d$ should satisfy Eq.~\eqref{eq:CritPointCond} with $k=0$. By substituting the decomposition \eqref{eq:DecompPRFIM} into the definition of the magnetization, we obtain the normalization condition for $f^{d}$:
\begin{equation}
1=z\int \dd h \,\text{sgn}(h)\,\left(f^d\,Q_c^MP_H^{crit}\right)(h).
\end{equation}
Let us call $f^{s}$ the left antisymmetric eigenvector of the longitudinal operator, and normalize it in such a way as $f^s\cdot f^{d}=1$. Expanding the cavity Eq.~\eqref{eq:RSsimplNotRFIM} close to criticality, and projecting on $f^s$ we obtain:
\begin{equation}
\label{eq:EqMagn}
\begin{split}
0=&\,M(M-1)\,m\,f^s\cdot\mathbf{R}_0\, f^d\,\delta S\,Q_c^{M-2}P_H^{crit}+\\
&+\frac{M(M-1)(M-2)}{6}m^3\,f^s\cdot\mathbf{R}_0\,(f^d)^3\,Q_c^{M-3}P_H^{crit}+\\
&+m\,M\,f^s\cdot \mathbf{R}_0f^dQ_c^{M-1}\delta P_H.
\end{split}
\end{equation}
Note that the terms in which there is a scalar product between $f^s$ and an even function vanish by symmetry, implying that there are no terms proportional to even powers of $m$.

For the Bethe lattice with connectivity equal to three there is no term due to the convolution $(f^d)^3$, and the equation can only be satisfied because of the term coming from the variation of the symmetric part $\delta S$. The expression for $\delta S$ can be obtained linearizing the cavity equation:
\begin{equation}
\begin{split}
\delta S = &\mathbf{R}_0\,Q_c^{M}\delta P_H+M\mathbf{R}_0Q_c^{M-1}(\delta S)P_H^{crit}+\\&+m^2\frac{M(M-1)}{2}\mathbf{R}_0(f^d)^2Q_c^{M-2}P_H^{crit}.
\end{split}
\end{equation}
At this point let us write $\delta S$ as the sum of two contributions:
\begin{equation}
\delta S=\delta_{\epsilon} S+\delta_m S,
\end{equation}
where $\delta_{\epsilon} S$ is the variation of $S$ with respect to $\epsilon$ at $m=0$, and $\delta_{m} S$ is the variation of $S$ with respect to $m$ at $\epsilon=0$. We have:
\begin{equation}
\label{eq:deltaSe}
\delta_{\epsilon}S=\mathbf{R}\,\delta_{\epsilon}S+\epsilon\,\zeta_{\epsilon},
\end{equation}
\begin{equation}
\label{eq:deltaSm}
\delta_m S = \mathbf{R}\,\delta_m S+m^2\zeta_m,
\end{equation}
where we defined
\begin{equation}
\begin{split}
\epsilon\,\zeta_{\epsilon}&=\mathbf{R}_0Q_c^{M}\delta P_H\\ \zeta_m &=\frac{M(M-1)}{2}\,\mathbf{R}_0\,(f^d)^2\,Q_c^{M-2}P_H^{crit}.
\end{split}
\end{equation}
Note that since the critical eigenvector is antisymmetric, $f^s\cdot\zeta_m=f^s\cdot\zeta_\epsilon=0$. Equations \eqref{eq:deltaSe} and \eqref{eq:deltaSm} can be inverted:
\begin{equation}
\delta_{\epsilon} S=\epsilon\,(\mathds{1}-\mathbf{R})^{-1}\zeta_{\epsilon},\quad \delta_{m} S=m^2(\mathds{1}-\mathbf{R})^{-1}\zeta_{m},
\end{equation}
allowing to write $\delta S$ in terms of $\zeta_{\epsilon}$ and $\zeta_m$. Going back to the original equation \eqref{eq:EqMagn} we obtain an expression of the form:
\begin{equation}
\label{eq:alpha}
-\alpha^2\,m\,\epsilon+m^3=0,
\end{equation}
where the coefficient $\alpha$ is given by:
\begin{widetext}
\begin{equation}
\label{eq:alphaComp}
\alpha=\left(-\frac{(M-1)f^s\cdot\mathbf{R}_0\big(f^d\left((\mathds{1}-\mathbf{R})^{-1}\zeta_{\epsilon}\right)Q_c^{M-1}P_H^{crit}\big)+f^s\cdot\xi_{\epsilon}}{(M-1)\,f^s\cdot\mathbf{R}_0\left(\frac{(M-2)}{6}(f^d)^3\,Q_c^{M-3}+f^d\big((\mathds{1}-\mathbf{R})^{-1}\zeta_m\big)Q_c^{M-1}\right)P_H^{crit}}\right)^{1/2},
\end{equation}
\end{widetext}
and $\xi_{\epsilon}$ is defined by:
\begin{equation}
\epsilon\,\xi_{\epsilon}=\mathbf{R}_0\,f^dQ_c^{M-1}\delta P_H.
\end{equation}
From \eqref{eq:alpha} it follows that close to the critical point the magnetization has a square root critical behavior:
\begin{equation}
m\approx\alpha\, \epsilon^{1/2}.
\end{equation}
It is interesting to note that the only difference between the case $T>0$ and $T=0$ is that in the latter $f^d(u)$, $Q_c(u)$, and $\delta S(u)$ have finite weight in $u=\pm 1$. The analysis based on the symmetries is the same, leading in both cases to $m=O(\sqrt{\epsilon})$. 
Expressions like Eq.~\eqref{eq:alphaComp} can be computed by discretizing $\mathbf{R}$ (Eq.~\eqref{eq:defGrassettoR}), i.e.\ by representing the distributions of the interval $[0,1]$ through a basis of histograms, and by computing the matrix elements associated with $\mathbf{R}$. In this way for $z=3$ we obtained 
\begin{equation}
\label{eq:valueAlphaZ3}
\alpha=3.71\dots
\end{equation}
See Fig.~\ref{eq:NfrozEquiv} for the comparison with the numerics. 

Interestingly the fact that the longitudinal operator develops a critical eigenvector at the same point of the integral equation associated with the susceptibility can be checked for $T>0$ by noticing that if $g(u)$ is the eigenvector of the susceptibility equation (Eq.~\eqref{eq:CritPointCond} with $k=1$), then $g'(u)$ is an eigenvector of the cavity equation (Eq.~\eqref{eq:CritPointCond} with $k=0$). This can be shown deriving the susceptibility equation, and performing an integration by parts. The terms coming from the boundaries $u=\pm 1$ vanish because $g(u)$ goes to zero continuously at $u=\pm 1$ at any {\it finite} temperature, due to the term proportional to the derivative of $\tilde{u}$ (see also Figs. $(2)$ and $(3)$ in Ref.~\cite{parisi2014diluted}). 

At $T=0$ also the susceptibility and the longitudinal operator should diverge at the same point. In this case the eigenvector of the susceptibility operator no longer vanishes at $u=\pm 1$ having instead a finite limit, see Fig.~\ref{fig:limitDistr}. Again there is a critical longitudinal anti-symmetric eigenvector that can be identified with $g'(u)$ for $|u|<1$. In addition however the eigenvector carries a finite weight at $|u|=1$ corresponding to the fact that the cavity equation has a finite weight at $|u|=1$ for $T=0$. The antisymmetric eigenvector carries a weight in $u=\pm 1$ that is exactly equal to $\mp g(1)$, as can be verified again by an integration by part, this time taking care of the fact that $g(\pm 1) \neq 0$. Formally one can say that the longitudinal eigenvector is proportional to $g'(u)$ with $g(u)$ that has a discontinuity from zero to $g(1)$ at $u=\pm 1$ that leads to the appearance of antisymmetric delta functions at the extrema.

\section{Computation of $p(\epsilon)$ close to the critical point}
\label{sec:CompQeps}
Once $m$ is know, it is possible to compute the critical exponent of the probability $p$ of drawing an open population. In order to do that let us express $n_{\text{unf}}$ in terms of $p$. Close to the critical point the total cavity marginal distribution on a site conditioned to the open populations is given by: 
\begin{equation}
P_o^{\text{site}}=z\,g^{\text{site}}_1 P(\Delta)+\dots,
\end{equation}
where by using the notation of Eq.~\eqref{eq:expancloscrit}, we defined 
\begin{equation}
g^{\text{site}}_1=g^d(Q^{crit})^{M}P_H^{crit}.
\end{equation}
We have
\begin{equation}
\label{eq:nflip}
\begin{split}
n_{\text{unf}}&=\int \dd h \,\dd \Delta\, P_o^{\text{site}}(h,\Delta_h)\,\mathds{1}\left(|h|<\Delta_h/2\right)=\\&=2z\int_0^{\infty}\dd h\, g^{\text{site}}_1(h) \int_{2\,h}^{\infty}\dd\Delta P(\Delta)+\dots\\&=z\, g^{\text{site}}_1(0)\frac{B}{B_2}p^2+\dots
\end{split}
\end{equation}
Note that only a fraction of order $p$ of the open populations contribute to the fraction of spins that change sign. Equation \eqref{eq:nflip} implies that
\begin{equation}
\label{eq:scalingQ}
p\approx\kappa \epsilon^{1/4},\quad  \kappa=\sqrt{\frac{\alpha\, B_2}{B z\,g^{\text{site}}_1(0)}},
\end{equation}
where $\kappa$, by using Eq.~\eqref{eq:valueAlphaZ3}, for $z=3$ equals
\begin{equation}
\kappa=1.90\dots
\end{equation}
where $B,B_2$ (see equations \eqref{eq:DefB} and \eqref{eq:DefB2}) and $g^{\text{site}}_1$ are computed in population dynamics. As shown in Fig.~\ref{fig:epsilon}, the numerical simulations are in perfect agreement with the analytical prediction. 

\section{The large $L$ limit of the chain in the paramagnetic phase}
\label{sec:LineParamagnPhase}
In this appendix we derive the properties discussed in section \ref{sec:ConnectionToCorrFunct} of the correlations between two points in the limit of large distances in the paramagnetic phase. 

Let us recall the definition of the $u$-extremes
\begin{equation}
\label{eq:extremesLine}
u^{\pm}_L=u_L\pm \frac{\Delta_L}{2},\quad \Delta_L=2|J_L|,
\end{equation}
that correspond to the maximum and minimum field acting on $\sigma_L$ when fixing $\sigma_0=\pm 1$ on the chain (see Eq.~\eqref{eq:twoSpinHamilt}). In order to deduce the iteration rule for the extremes let us construct a chain of length $L+1$, starting from a chain of length $L$. The first step is to draw a field $h_{iter}$ to put on $\sigma_L$, where
\begin{equation}
h_{iter}=H+\sum_{i=1}^{M-1}u_i,
\end{equation}
and
\begin{equation}
H\sim P_H,\quad u_1,\ldots,u_{M-1}\,\overset{\text{i.i.d.}}{\sim}\,Q_{RS}(u).
\end{equation}
Next we add a new spin $\sigma_{L+1}$ connected to $\sigma_L$ with a new coupling $J$, drawn from $P_J$. In the end, since we are at zero temperature, we optimize the energy over $\sigma_L$:
\begin{multline}
\label{eq:ricorsioneLinea}
\mathcal{H}_{L+1}(\sigma_0,\sigma_{L+1})=\\=\min_{\sigma_L}\big[\mathcal{H}_{L}(\sigma_0,\sigma_{L})-\sigma_L(h_{iter}+J\,\sigma_{L+1})\big]=\\= E-u_0'\sigma_0-\sigma_0\,J_{L+1}\,\sigma_{L+1}-u_{L+1}\sigma_{L+1}.
\end{multline}
Equation \eqref{eq:ricorsioneLinea} defines the updating rule linking $(u_0,J_L,u_L)$ with $(u_0',J_{L+1},u_{L+1})$. If we focus on the couple of fields acting on the spin at distance $L$ we obtain
\begin{equation}
\label{eq:iterExtrchain}
u_{L+1}^{+}=f^{(+)}_{J}(h^{+}_L,h^{-}_L),\quad u_{L+1}^{-}=f^{(-)}_{J}(h^{+}_L,h^{-}_L),
\end{equation}
\begin{equation}
\label{eq:FromUtoH}
h^{\pm}_L=h_{iter}+u^{\pm}_L,\quad h_{iter}=H+\sum_{i=1}^{M-1}u_i,
\end{equation}
where $f^{(+)}$ and $f^{(-)}$ are the same ordering functions \eqref{eq:ordFunc} that we used for the equations of the RSB extremes. As we already argued in section \ref{sec:ConnectionToCorrFunct}, Eqs.~\eqref{eq:FromUtoH} and \eqref{eq:iterExtrchain} are formally analogous to the equation for the extremes in the case of a single RSB population.

In order to study the statistical properties of the extremes acting on $\sigma_L$, let us consider the joint probability distributions $Q_o^{(L)}(u_L,\Delta_L)$ and $P_o^{(L)}(h_L,\Delta_L)$ of the open couples  on a chain of length $L$. From \eqref{eq:FromUtoH} and \eqref{eq:iterExtrchain} we obtain
\begin{align}
\label{eq:QoLine}
& Q_o^{(L+1)} = \mathbf{F}_1 P_o^{(L)}, \\
& P_o^{(L)} = Q_o^{(L)}\,Q_{RS}^{M-1}P_H,
\end{align}
where $\mathbf{F}_1$ is the same operator we defined in \eqref{eq:defF1} for the RSB extremes. Note that in the RSB case the limit of small width of the extremes is valid when approaching the critical point. Here, since there is only a single open couple, 
the effective coupling cannot increase during the iteration, i.e.\ $|J_{L+1}|\leq|J_L|$, and then, independently of the distance from the transition, one expects the typical $J_L$ to be small for large $L$. Then for studying the large $L$ behavior of \eqref{eq:QoLine} we can rely on the same expansion we did in appendix \ref{sec:ExpansionCloseToCrit}. As for the RSB extremes we define:
\begin{equation}
g_L(s)=g^s\cdot q_o^{(L)}(s),
\end{equation}
where $q_o^{(L)}(s)$ is the Laplace transform of $Q_o^{(L)}$ and $g^s$, $g^d$ are the left and right eigenvectors associated with the maximum eigenvalue $\lambda$ of 
\beq
\label{eq:F0Chain}
\mathbf{F}_0 Q_{RS}^{M-1}P_H,
\eeq
for an arbitrary $\sigma_H>\hat \sigma_H$.
Following the same steps leading to \eqref{eq:ScalinoRSB}, we find:
\begin{multline}
\label{eq:ScalinoChain}
g^s\cdot(\mathbf{L}\mathbf{\Delta F_1} P_o^{(L)})(s)=\\=g^s(1)g_1(1)\left(2\,\frac{g_L(s)-g_L(0)}{s} -\dot{g}_L(s)\right),
\end{multline}
where in analogy with the RSB case we used the notation:
\begin{equation}
g_1=g^dQ_{RS}^{M-1}P_H.
\end{equation}
Projecting the Laplace transform of \eqref{eq:QoLine} on $g^s$, we find
\begin{multline}
\label{eq:ExpansionChain}
g_{L+1}(s)=\lambda \,g_{L}(s)+\\+g^s(1)g_1(1)\left( 2\,\frac{g_{L}(s)-g_{L}(0)}{s} -\dot{g}_{L}(s)\right).
\end{multline}
We stress that here the expansion only requires $L$ to be large, while $\sigma_H$ can be arbitrarily larger than $\hat{\sigma}_H$. For large $L$, Eq.~\eqref{eq:ExpansionChain} admits a solution of the form
\begin{equation}
\label{eq:AnsatzChain}
g_L(s) \approx \lambda^L L \,\varphi(s/L),   
\end{equation}
where $\varphi$ is to be determined. Substituting \eqref{eq:AnsatzChain} in \eqref{eq:ExpansionChain}, and expanding for large $L$, we have
\begin{multline}
g_{L+1}(s)-\lambda \, g_{L}(s) \approx \lambda^{L+1} \varphi(s/L)+\\- \lambda^{L+1}s/L\, \frac{d}{d\,(s/L)}\varphi(s/L),
\end{multline}
and
\begin{multline}
2 \frac{g_L(s)-g_L(0)}{s}- \dot{g}_L(s) \approx\\\approx \lambda^L \,  \left(2\frac{\varphi(s/L)-\varphi(0)}{s/L}-\frac{d}{d\,(s/L)}\varphi(s/L)\right),     
\end{multline}
from which we obtain the following equation for $\varphi$:
\begin{multline}
\varphi(s/L)-s/L\,\frac{d}{d\,(s/L)}\varphi(s/L)=\\=\Gamma\left(2\frac{\varphi(s/L)-\varphi(0)}{s/L}-\frac{d}{d\,(s/L)}\varphi(s/L)\right), 
\end{multline}
where we defined the constant
\begin{equation}
\Gamma=\frac{1}{\lambda}g^s(1)\,g_1(1),
\end{equation}
that at the critical point is equal to $B_2$ (see Eq.~\eqref{eq:DefB2}).
By changing variables in order to set all the constants to one:
\begin{equation}
z=\frac{s}{\Gamma\,L},\quad y(z)=\frac{\varphi(s/L)}{\varphi(0)},
\end{equation}
one finds 
\begin{equation}
y(z)-z\,\dot{y}(z)=\left(2\frac{y(z)-1}{z}-\dot{y}(z)\right), 
\end{equation}
that has solution
\begin{equation}
\label{eq:SolChain}
y(z)=\frac{1+c\,z^2}{1-z},
\end{equation}
with $c$ an undetermined constant. The inverse transform of \eqref{eq:SolChain} is
\begin{equation}
\label{eq:Solutionf}
f(\Delta)=e^{-\Delta}+c\,\left(e^{-\Delta}-\delta(\Delta)+\delta'(\Delta)\right), 
\end{equation}
from which we argue that $c=0$, since the singular terms in \eqref{eq:Solutionf} can only result from a non physical initialization of $Q_o^{(1)}$. Therefore for large $L$ the probability $q_L$ of drawing a couple of spins with non-zero effective coupling is (see \eqref{eq:AnsatzChain})
\begin{equation}
q^{(J)}_L=L\lambda^L,
\end{equation}
and the distribution of the open effective coupling is an exponential, with mean value that scales linearly in $1/L$:
\begin{equation}
\label{eq:AveChain}
L\,\big\langle |J|\big\rangle_{J>0}=\frac{1}{2\,\Gamma(\sigma_H) }\xrightarrow[]{\sigma_H\downarrow \hat{\sigma}_H}\frac{1}{2B_2}=1.326...
\end{equation}
This behavior, and in particular the value of the average coupling at the critical point are in perfect agreement with the interpolations obtained from numerical data in Ref.~\cite{angelini2019new}. Note that all the quantities that we are considering here remain regular at the transition, however for the theory to be consistent the computation of the critical point should lead to the same results from both sides of the dAT line. This is guaranteed from the fact that for $\sigma_H=\hat{\sigma}_H$
the maximum eigenvalue of \eqref{eq:F0Chain} becomes $\lambda=1/M$, and the spin glass susceptibility diverges because of the summation over all the pairs of spins (see Ref.~\cite{parisi2014diluted} for all the definitions and details).

\section{The Ginzburg criterion from the RSB phase}
\label{sec:GinzSGphase}
In this section we study the Ginzburg criterion in the spin-glass phase. The strategy is to compute the fluctuations of a suitably chosen order parameter at the leading order in $1/M$, and to check at which dimension they become important.

\subsection{Percolation}
It is instructive to first consider the percolation problem. Let us call $\tau$ the occupancy probability of a node. Let us call $\mathcal{P}$ the percolating cluster, i.e.\ the set of all occupied sites whose elements have at least a neighbour in $\mathcal{P}$. Given a node $i$ on the Bethe lattice consider the cavity graph obtained removing an edge connected to $i$. The probability $p$ that $i$ belongs to the percolating cluster on such cavity graph can be computed self-consistently according to the following equation:
\begin{equation}
\label{eq:cavEqPerc}
1-p=(1-\tau)+\tau\,(1-p)^{z-1}.
\end{equation}
Once \eqref{eq:cavEqPerc} is solved, it is possible to compute the probability $p_n$ that a node on the original graph belongs to $\mathcal{P}$:
\begin{equation}
p_n=\tau\,\big(1-(1-p)^z\big).
\end{equation}
Equation \eqref{eq:cavEqPerc} develops a solution with $p \neq 0$ for $\tau<\tau_c \equiv 1/(z-1)$, where $\tau_c$ is the critical occupancy probability. In particular, in the proximity of $\tau_c$:
\begin{equation}
p=\frac{2}{z-2}\, \epsilon+O(\epsilon^2),\quad \epsilon=(z-1)\,\tau-1.
\end{equation}
At this point let us consider a chain composed by $L$ edges. The fluctuations of the order parameter are given by 
\begin{equation}
\label{eq:ConnCorrPerc}
C^{\text{\tiny BL}}(L)=q^{\text{\tiny BL}}(L)-p_n^2,
\end{equation}
where $q^{\text{\tiny BL}}(L)$ is the probability that two sites at distance $L$ belong to the percolating cluster. Close to the critical point the probability that a node belongs to $\mathcal{P}$ is small, and therefore $q^{\text{\tiny BL}}(L)$ can be expanded as follows:
\begin{equation}
\label{eq:Expansionq0L}
q^{\text{\tiny BL}}(L)=q^{(1)}(L)+q^{(2)}(L)+O(p^3),
\end{equation}
where $q^{(n)}(L)$ is the probability that the two ends of the chain belong to $\mathcal{P}$ because of the presence of $n$ percolating nodes connected to the chain. Let us call ``source node'' a node belonging to the chain that is connected to one of such percolating nodes. Note that for the computation of \eqref{eq:ConnCorrPerc} we only need $q^{\text{\tiny BL}}(L)$ up to order $\tau^2$. For large $L$ the leading contribution is
\begin{equation}
q^{(1)}(L)\approx p\,L\,\lambda^{L},
\end{equation}
where the factor $L$ comes from the summation over all possible positions of the source node, and $\lambda$ is the probability:
\begin{equation}
\lambda=\tau\,(1-p)^{z-2}.
\end{equation}
At this point let us study the correction $q^{(2)}(L)$. Let us number the nodes of the chain starting from one of the two ends, and let us identify the two source nodes respectively by $k_1$ and $k_2$, with $k_1\leq k_2$. If we use the notation $L_1=k_1$, and $L_2=L-k_2$ we can write
\begin{equation}
\label{eq:q20L}
q^{(2)}(L)= \sum_{1,2}\,\tilde{p}(L_1)\,\tilde{p}(L_2)+O\left(p^2\,L\,\lambda^L\right),
\end{equation}
where the sum 
\begin{equation}
\sum_{1,2}\equiv\sum_{L_1=0}^{L-1}\sum_{I=1}^L\sum_{L_2=0}^{L-1}\delta(L-L_1-I-L_2)
\end{equation}
is over all disjoint couples $k_1\neq k_2$, and $\tilde{p}(R)$ is the probability that a chain of length $R$ having a source node in one of its ends belongs to $\mathcal{P}$. The probability $\tilde{p}(R)$ is given by:
\begin{equation}
\tilde{p}(R)=p\,(1-\lambda)\,(1-p)\,\lambda^{L_1}\left(1-\delta(R)\right)+p\,\delta(R),
\end{equation}
where we used the identity
\begin{equation}
p\,(1-\lambda)=\tau\,\big(1-(1-p)^{z-2}\big).
\end{equation}
The last term in \eqref{eq:q20L} is due to the case in which $k_1=k_2$, and it is not written explicitly because it does not contribute at the leading order to $C^{\text{\tiny BL}}$. By summing all the terms
\begin{equation}
\sum_{1,2}\left(1-\delta(L_1)\right)\,\delta(L_2)\,\lambda^{L_1}=\sum_{L_1=1}^{L-1}\lambda^{L_1}=\frac{\lambda}{1-\lambda}-\frac{\lambda^{L}}{1-\lambda},
\end{equation}
\begin{multline}
\sum_{1,2}\left(1-\delta(L_1)\right)\,\left(1-\delta(L_2)\right)\,\lambda^{L_1+L_2}=\\=\sum_{I=1}^{L-2}(L-I-1)\,\lambda^{L-I}\approx \frac{\lambda^2}{(1-\lambda^2)}-\frac{L\,\lambda^L}{1-\lambda},
\end{multline}
we find
\begin{equation}
q^{(2)}(L)=p_n^2+O\left(p^2L\lambda^L\right),
\end{equation}
and therefore
\begin{equation}
\label{eq:PercolGBL}
C^{\text{\tiny BL}}(L)\propto p\,L\,\lambda^{L}+O\left(p^2L\lambda^L\right).
\end{equation}
At this point we can substitute \eqref{eq:combinTermMlayer} and \eqref{eq:PercolGBL} into \eqref{eq:leadingMcorr}, and compare the two point function, computed on the scale of the correlation length $\xi=O(\epsilon^{-1/2})$, with the square probability of belonging to the percolating cluster: 
\begin{equation}
\label{eq:GinzburgPercolation}
\frac{1}{p_n^2}C(b\,\xi)\propto\frac{\epsilon^{d/2-3}}{M}\int_0^{\infty}\frac{\dd\alpha}{\alpha^{d/2-1}}\exp{\left(-b^2/(4\alpha)-\alpha\right)},
\end{equation}
that is the so called Ginzburg parameter. In \eqref{eq:GinzburgPercolation} we used 
\begin{equation}
-\ln \lambda (z-1)\approx  -\ln (1 - \epsilon)  \approx \epsilon.
\end{equation}
Note that at a fixed distance from the critical point the correction is small provided $M$ is large. This corresponds to the regime in which the correlation length is smaller than the typical tree-like neighbourhood of a site. However for fixed $M$, as soon as $d<6$ the Ginzburg parameter diverges, in agreement with the expected result $D_U=6$ for the percolation problem (see Ref.~\cite{stauffer2018introduction}). 

\subsection{The spin glass}
\begin{figure}
\centering
\includegraphics[width=\columnwidth]{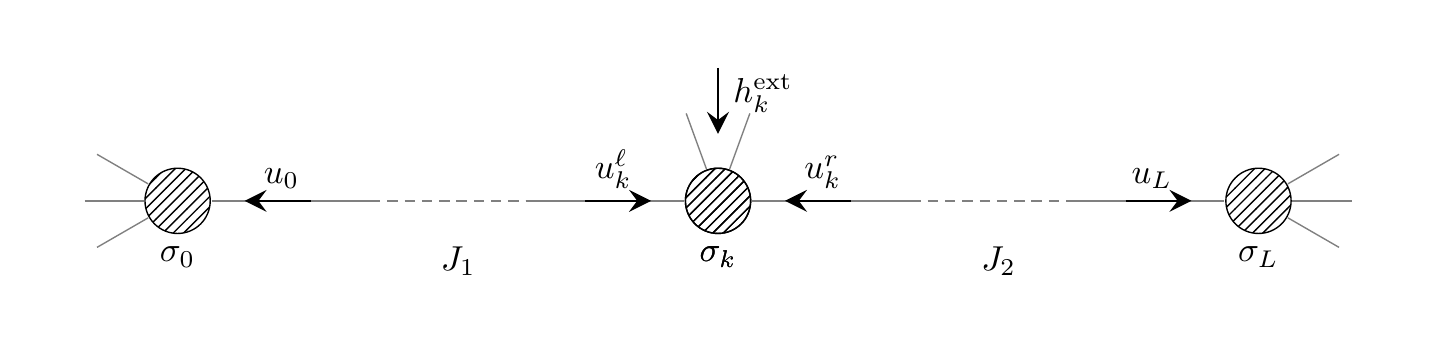}
\caption{The leading contribution to the open-open correlation function close to the critical point is given by a chain in which the extremal sites $\sigma_0$ and $\sigma_L$ are open because of an open population $h^{\text{ext}}_k$ entering $\sigma_k$.\label{fig:ChainGinzburg} 
} 
\end{figure}
At this point let us consider the SG below the dAT line. As we have seen, in the RSB phase we can define a local order parameter corresponding to the indicator function of the open sites, that is equal to one if there is at least an open couple entering the site, and zero otherwise. The average order parameter is thus $p_{\text{rsb}}$, and its fluctuations are given by the probability that two sites at distance $x$ are both open. Let us compute these fluctuations on the Bethe lattice.

Analogously to the percolation case, also for the SG we can write Eq.~\eqref{eq:Expansionq0L}. Here $q^{(n)}(L)$ represents the probability that both ends $0$ and $L$ of the chain are open because of the presence of $n$ open spins connected to the chain. The leading contribution $q^{(1)}(L)$ can be obtained by taking two triplets $(u_0,J_1,u_k^{\ell})$ and $(u_k^r,J_2,u_L)$ in which all the fields are closed, and by joining them with the insertion of an open field $h^{\text{ext}}_k$ (see Fig.~\ref{fig:ChainGinzburg}). In particular
\begin{equation}
\label{eq:hExtk}
h^{\text{ext}}_k=H+\sum_{i=1}^{z-3}u_i+u_o,
\end{equation}
where the $u_i$'s are closed and $u_o$ is an open population. Let us use as usual the notation $\Delta=u_o^+-u_o^-$. At this point for $0$ and $L$ to be open, the two chains must necessarily have $J_1 \neq 0$ and $J_2 \neq 0$ (see appendix \ref{sec:LineParamagnPhase}). This is also a sufficient condition if the central spin $\sigma_k$ is not frozen, i.e.\ if it has not always the same magnetization over all the local ground states (LGS). Indeed in this case the flipping of $\sigma_k$ leads to a change in the fields acting on the ends of the chain $\sigma_0,\sigma_L$. If $\sigma_0$ and $\sigma_L$ receive local fields that do not have the same values over all the LGS they are open by definition. Therefore we have:  
\begin{multline}
\label{eq:CorrOpenOpen}
q^{(1)}(L)\approx\sum_{k=0}^L \Big\langle P_c(h^{\text{ext}}_0)\,P_o^{(k)}(u_0,J_1,u_{k}^{\ell})\,\widetilde{P}_o(h^{\text{ext}}_{k},\Delta)\times\\\times\mathds{1}(|h_{k}|<\Delta/2)\, P_o^{(L-k)}(u_{k}^r,J_2,u_L)\,P_c(h^{\text{ext}}_L)\Big\rangle,
\end{multline}
where the angle brackets represent an average over all the arguments, $h_k$ is the total average field acting on $\sigma_k$:
\begin{equation}
\label{eq:defhk}
h_{k}=\hat{u}(J_1,h^{\text{ext}}_0+u_0)+u_{k}^{\ell}+h^{\text{ext}}_{k}+u_{k}^r+\hat{u}(J_2,u_L+h^{\text{ext}}_L),
\end{equation} 
$\widetilde{P}_o$ is given by
\begin{equation}
\widetilde{P}_o=(z-2)\,Q_o(Q_c)^{z-3}P_H,
\end{equation}
and $P_o^{(L)}(u,J,u')$ is the probability distribution of the triplet $(u,J,u')$ conditioned to $J\neq 0$. Note that in \eqref{eq:CorrOpenOpen} the indicator function enforces the condition that $h_k+\Delta/2>0$ and $h_k-\Delta/2<0$, that is equivalent to require that $\sigma_k$ is not frozen. Note also that in Eq.~\eqref{eq:defhk} the terms in $\hat{u}$ are obtained from the minimization of $\sigma_0$ and $\sigma_L$. This is a consequence of the fact that at the leading order the local field on the extremal spins is open, but $\sigma_0$ and $\sigma_L$ assume the same value (the value obtained with the minimization) on all the states.

For large lengths $L_1$ and $L_2$ of the chains we can use the approximation (see section \ref{sec:ConnectionToCorrFunct} and appendix \ref{sec:LineParamagnPhase}):
\beq
\label{eq:longChainapprox}
P_o^{(L)}(u,J,u')\propto L \, \lambda^L g^d(u)\,g^d(u')\frac{1}{2\,J_{typ}}e^{-|J|/J_{typ}},
\eeq 
where $J_{typ}=1/(2B_2L)$, and all functions are those corresponding to the situation above the dAT line, except for $\lambda$, that is such that $(z-1) \lambda$ is smaller than one, because the external populations are closed. More precisely we have:
\begin{equation}
\label{eq:shiftAutoval}
(z-1)\lambda=(z-1)\,g^s\cdot\mathbf{F}_0g^dQ_c^{M-1}P_H,
\end{equation}
that can be expanded close to criticality remembering that the shift of the distribution of the closed populations obeys (see Eq.~\eqref{eq:variationClosedPop}):
\beq
\delta Q_c\approx - p\, g^d.
\eeq 
In this way we obtain 
\begin{equation}
(z-1)\lambda\approx 1-2Bp,
\end{equation}
where the constant $B$ is defined in Eq.~\eqref{eq:DefB}. Note that we considered only the variation with respect to $Q_c$, since the two triplets before the insertion of the open population belong separately to the cluster of closed spins.  
At this point since for large $L_1$ and $L_2$ the couplings $J_1$ and $J_2$ are small, at the leading order we can also neglect in \eqref{eq:defhk} the contributions coming from $\sigma_0$ and $\sigma_L$:
\begin{equation}
h_k\approx u_k^{\ell}+h^{\text{ext}}_k+u_k^r.
\end{equation}
These considerations allow us to perform in \eqref{eq:CorrOpenOpen} the integrations over $J_1$, $J_2$, the fields, and the summations over $L_1$ with $L_2= L-L_1$, obtaining:
\begin{equation}
q^{(1)}(L)\approx p\,\Delta_{typ}\lambda^{L}\sum_{L_1=0}^L L_1 \, (L-L_1)\approx p\,\Delta_{typ}\,\lambda^{L}L^3.
\end{equation}
For the term $q^{(2)}(L)$ we have to consider the insertion of two open populations on two spins $\sigma_{k_1},\sigma_{k_2}$ of the chain. This can be done by joining three triplets, $(u_0,J_1,u_1^{\ell})$, $(u^r_1,J_I,u^{\ell}_2)$ and $(u_2^r,J_2,u_L)$. Again for the extremes to be open, $\sigma_{k_1}$ and $\sigma_{k_2}$ should be not frozen, and $J_1,J_2\neq 0$. Therefore
defining $L_1=k_1$, and  $L_2=L-k_2$, we have:
\begin{widetext}
\begin{equation}
\label{eq:q0L2}
\begin{split}
q^{(2)}(L)=&\sum_{1,2}\bigg\langle  P_c(h^{\text{ext}}_0)\,P_o^{(L_1)}(u_0,J_1,u_1^{\ell})\,\widetilde{P}_o(h^{\text{ext}}_1,\Delta_1)\,\mathds{1}(|h_1|<\Delta_1/2)\,P^{(I)}(u^r_1,J_I,u^{\ell}_2)\times\\
&\hphantom{\sum_{1,2}\bigg\langle}\times\mathds{1}(|h_2|<\Delta_2/2)\,\widetilde{P}_o(h^{\text{ext}}_2,\Delta_2)\,P_o^{(L_2)}(u_2^r,J_2,u_L)\, P_c(h^{\text{ext}}_L)\bigg\rangle+O\left(p^2\,\Delta_{typ}^2, \,L^3\lambda^L\right)
\end{split}
\end{equation}
\end{widetext}
where the subscripts $1,2$ of the fields refer respectively to $k_1$ and $k_2$, and $P^{(I)}$ is the joint probability distribution of $(u^r_1,J_I,u^{\ell}_2)$. Analogously to the percolation case the last term in \eqref{eq:q0L2}, that we did not write explicitly, is due to the case in which $k_1=k_2$. Note that here the correlation between $0$ and $L$ is not only due to the constraint $L=L_1+I+L_2$, like in the percolation case, but in general it also depends on the central triplet $(u^r_1,J_I,u^{\ell}_2)$. Since we are interested in the large length limit, we can use the asymptotic formula:
\begin{multline}
\label{eq:PIasympt}
P^{(I)}(u_1,J_I,u_2) \approx Q_c(u_1) \, Q_c(u_2)\,\delta(J_I)+\\ + a_1\, L \, \lambda^L g^d(u_1)\, g^d(u_2) \, \left( \frac{1}{2 \, J_{typ}} \, e^{- |J_I|/J_{typ}} -\delta(J_I)  \right),
\end{multline}
that is the same as Eq.~\eqref{eq:longChainapprox} (see appendix \ref{sec:LineParamagnPhase}) except for another contribution coming from the chain with $J_I=0$ (see Refs.~\cite{angelini2020loop,angelini2022unexpected}). This new term does not contribute to $q^{(1)}$, since we had to require $J_1,J_2\neq 0$, while it is fundamental to take into account in this case. When averaging over $J_I$, \eqref{eq:PIasympt} becomes
\begin{equation}
\int\dd J_I P_I(u_1,J_I,u_2)\approx Q_c(u_1) \, Q_c(u_2),
\end{equation}
and then, at the leading order, the two total fields $h_1$ and $h_2$ are independent. To be more precise the expression \eqref{eq:q0L2} for $q^{(2)}(L)$ becomes completely equivalent to the percolation case. In particular one can use the same notation of \eqref{eq:q20L} with the definition 
\begin{equation}
\begin{split}
\tilde{p}(L_1)=& \bigg\langle  P_c(h_o^{\text{ext}})\,P_o^{(L_1)}(u_0,J_1,u_{1}^{\ell})\,\widetilde{P}_o(h^{\text{ext}}_{1},\Delta_{1})\times\\ &\hphantom{\bigg\langle}\,\times\mathds{1}(|h_{1}|<\Delta_{1}/2)\,Q_c(u_1^r)\bigg\rangle \left(1-\delta(L_1)\right)+\\&+p\,\delta(L_1),
\end{split}
\end{equation}
where the first term in angular brackets is proportional to
\begin{equation}
p\,\Delta_{typ} L\lambda^L.
\end{equation}
Following the same steps of percolation, and taking into account that close to the critical point $p\propto \pRSB$ and that $\Delta_{typ}\propto \pRSB$, we find the fluctuation of the order parameter written in Eq.~\eqref{eq:FluctOrdParBLmain}.
\clearpage
\bibliography{bibliografia.bib}

\begin{thebibliography}{54}%
\makeatletter
\providecommand \@ifxundefined [1]{%
 \@ifx{#1\undefined}
}%
\providecommand \@ifnum [1]{%
 \ifnum #1\expandafter \@firstoftwo
 \else \expandafter \@secondoftwo
 \fi
}%
\providecommand \@ifx [1]{%
 \ifx #1\expandafter \@firstoftwo
 \else \expandafter \@secondoftwo
 \fi
}%
\providecommand \natexlab [1]{#1}%
\providecommand \enquote  [1]{``#1''}%
\providecommand \bibnamefont  [1]{#1}%
\providecommand \bibfnamefont [1]{#1}%
\providecommand \citenamefont [1]{#1}%
\providecommand \href@noop [0]{\@secondoftwo}%
\providecommand \href [0]{\begingroup \@sanitize@url \@href}%
\providecommand \@href[1]{\@@startlink{#1}\@@href}%
\providecommand \@@href[1]{\endgroup#1\@@endlink}%
\providecommand \@sanitize@url [0]{\catcode `\\12\catcode `\$12\catcode
  `\&12\catcode `\#12\catcode `\^12\catcode `\_12\catcode `\%12\relax}%
\providecommand \@@startlink[1]{}%
\providecommand \@@endlink[0]{}%
\providecommand \url  [0]{\begingroup\@sanitize@url \@url }%
\providecommand \@url [1]{\endgroup\@href {#1}{\urlprefix }}%
\providecommand \urlprefix  [0]{URL }%
\providecommand \Eprint [0]{\href }%
\providecommand \doibase [0]{https://doi.org/}%
\providecommand \selectlanguage [0]{\@gobble}%
\providecommand \bibinfo  [0]{\@secondoftwo}%
\providecommand \bibfield  [0]{\@secondoftwo}%
\providecommand \translation [1]{[#1]}%
\providecommand \BibitemOpen [0]{}%
\providecommand \bibitemStop [0]{}%
\providecommand \bibitemNoStop [0]{.\EOS\space}%
\providecommand \EOS [0]{\spacefactor3000\relax}%
\providecommand \BibitemShut  [1]{\csname bibitem#1\endcsname}%
\let\auto@bib@innerbib\@empty
\bibitem [{\citenamefont {Sherrington}\ and\ \citenamefont
  {Kirkpatrick}(1975)}]{SK}%
  \BibitemOpen
  \bibfield  {author} {\bibinfo {author} {\bibfnamefont {D.}~\bibnamefont
  {Sherrington}}\ and\ \bibinfo {author} {\bibfnamefont {S.}~\bibnamefont
  {Kirkpatrick}},\ }\href@noop {} {\bibfield  {journal} {\bibinfo  {journal}
  {Phys.\ Rev.\ Lett.}\ }\textbf {\bibinfo {volume} {35}},\ \bibinfo {pages}
  {1792} (\bibinfo {year} {1975})}\BibitemShut {NoStop}%
\bibitem [{\citenamefont {Parisi}(1979)}]{Parisi1979}%
  \BibitemOpen
  \bibfield  {author} {\bibinfo {author} {\bibfnamefont {G.}~\bibnamefont
  {Parisi}},\ }\href@noop {} {\bibfield  {journal} {\bibinfo  {journal} {Phys.
  Rev. Lett.}\ }\textbf {\bibinfo {volume} {43}},\ \bibinfo {pages} {1754}
  (\bibinfo {year} {1979})}\BibitemShut {NoStop}%
\bibitem [{\citenamefont {Parisi}(1980{\natexlab{a}})}]{parisi1980order}%
  \BibitemOpen
  \bibfield  {author} {\bibinfo {author} {\bibfnamefont {G.}~\bibnamefont
  {Parisi}},\ }\href@noop {} {\bibfield  {journal} {\bibinfo  {journal} {J.\
  Phys.\ A}\ }\textbf {\bibinfo {volume} {13}},\ \bibinfo {pages} {1101}
  (\bibinfo {year} {1980}{\natexlab{a}})}\BibitemShut {NoStop}%
\bibitem [{\citenamefont {Parisi}(1980{\natexlab{b}})}]{parisi1980sequence}%
  \BibitemOpen
  \bibfield  {author} {\bibinfo {author} {\bibfnamefont {G.}~\bibnamefont
  {Parisi}},\ }\href@noop {} {\bibfield  {journal} {\bibinfo  {journal} {J.\
  Phys.\ A}\ }\textbf {\bibinfo {volume} {13}},\ \bibinfo {pages} {L115}
  (\bibinfo {year} {1980}{\natexlab{b}})}\BibitemShut {NoStop}%
\bibitem [{\citenamefont {de~Almeida}\ and\ \citenamefont
  {Thouless}(1978)}]{DeAlmeida1978}%
  \BibitemOpen
  \bibfield  {author} {\bibinfo {author} {\bibfnamefont {J.~R.}\ \bibnamefont
  {de~Almeida}}\ and\ \bibinfo {author} {\bibfnamefont {D.~J.}\ \bibnamefont
  {Thouless}},\ }\href {https://doi.org/10.1088/0305-4470/11/5/028} {\bibfield
  {journal} {\bibinfo  {journal} {Journal of Physics A: Mathematical and
  General}\ }\textbf {\bibinfo {volume} {11}},\ \bibinfo {pages} {983}
  (\bibinfo {year} {1978})}\BibitemShut {NoStop}%
\bibitem [{\citenamefont {M\'{e}zard}\ and\ \citenamefont
  {Parisi}(2001)}]{Mezard2001}%
  \BibitemOpen
  \bibfield  {author} {\bibinfo {author} {\bibfnamefont {M.}~\bibnamefont
  {M\'{e}zard}}\ and\ \bibinfo {author} {\bibfnamefont {G.}~\bibnamefont
  {Parisi}},\ }\href {https://doi.org/10.1007/PL00011099} {\bibfield  {journal}
  {\bibinfo  {journal} {Eur. Phys. J. B}\ }\textbf {\bibinfo {volume} {20}},\
  \bibinfo {pages} {217} (\bibinfo {year} {2001})}\BibitemShut {NoStop}%
\bibitem [{\citenamefont {M\'{e}zard}\ and\ \citenamefont
  {Parisi}(2003)}]{Mezard2003}%
  \BibitemOpen
  \bibfield  {author} {\bibinfo {author} {\bibfnamefont {M.}~\bibnamefont
  {M\'{e}zard}}\ and\ \bibinfo {author} {\bibfnamefont {G.}~\bibnamefont
  {Parisi}},\ }\href {https://doi.org/10.1023/A:1022221005097} {\bibfield
  {journal} {\bibinfo  {journal} {J. Stat. Phys.}\ }\textbf {\bibinfo {volume}
  {111}},\ \bibinfo {pages} {1} (\bibinfo {year} {2003})}\BibitemShut {NoStop}%
\bibitem [{\citenamefont {Panchenko}(2016)}]{panchenko2016structure}%
  \BibitemOpen
  \bibfield  {author} {\bibinfo {author} {\bibfnamefont {D.}~\bibnamefont
  {Panchenko}},\ }\href@noop {} {\bibfield  {journal} {\bibinfo  {journal}
  {Journal of statistical physics}\ }\textbf {\bibinfo {volume} {162}},\
  \bibinfo {pages} {1} (\bibinfo {year} {2016})}\BibitemShut {NoStop}%
\bibitem [{\citenamefont {Parisi}(2017)}]{parisi2017marginally}%
  \BibitemOpen
  \bibfield  {author} {\bibinfo {author} {\bibfnamefont {G.}~\bibnamefont
  {Parisi}},\ }\href@noop {} {\bibfield  {journal} {\bibinfo  {journal}
  {Journal of Statistical Physics}\ }\textbf {\bibinfo {volume} {167}},\
  \bibinfo {pages} {515} (\bibinfo {year} {2017})}\BibitemShut {NoStop}%
\bibitem [{\citenamefont {Concetti}(2019)}]{concetti2019properties}%
  \BibitemOpen
  \bibfield  {author} {\bibinfo {author} {\bibfnamefont {F.}~\bibnamefont
  {Concetti}},\ }\href@noop {} {\bibfield  {journal} {\bibinfo  {journal}
  {arXiv preprint arXiv:1908.03820}\ } (\bibinfo {year} {2019})}\BibitemShut
  {NoStop}%
\bibitem [{\citenamefont {De~Santis}\ and\ \citenamefont
  {Parisi}(2018)}]{de2018computation}%
  \BibitemOpen
  \bibfield  {author} {\bibinfo {author} {\bibfnamefont {F.}~\bibnamefont
  {De~Santis}}\ and\ \bibinfo {author} {\bibfnamefont {G.}~\bibnamefont
  {Parisi}},\ }\href@noop {} {\bibfield  {journal} {\bibinfo  {journal} {arXiv
  preprint arXiv:1805.01228}\ } (\bibinfo {year} {2018})}\BibitemShut {NoStop}%
\bibitem [{\citenamefont {Goldschmidt}\ and\ \citenamefont
  {Lai}(1990)}]{goldschmidt1990finite}%
  \BibitemOpen
  \bibfield  {author} {\bibinfo {author} {\bibfnamefont {Y.}~\bibnamefont
  {Goldschmidt}}\ and\ \bibinfo {author} {\bibfnamefont {P.-Y.}\ \bibnamefont
  {Lai}},\ }\href@noop {} {\bibfield  {journal} {\bibinfo  {journal} {Journal
  of Physics A: Mathematical and General}\ }\textbf {\bibinfo {volume} {23}},\
  \bibinfo {pages} {L775} (\bibinfo {year} {1990})}\BibitemShut {NoStop}%
\bibitem [{\citenamefont {De~Dominicis}\ and\ \citenamefont
  {Goldschmidt}(1989)}]{de1989replica}%
  \BibitemOpen
  \bibfield  {author} {\bibinfo {author} {\bibfnamefont {C.}~\bibnamefont
  {De~Dominicis}}\ and\ \bibinfo {author} {\bibfnamefont {Y.}~\bibnamefont
  {Goldschmidt}},\ }\href@noop {} {\bibfield  {journal} {\bibinfo  {journal}
  {Journal of Physics A: Mathematical and General}\ }\textbf {\bibinfo {volume}
  {22}},\ \bibinfo {pages} {L775} (\bibinfo {year} {1989})}\BibitemShut
  {NoStop}%
\bibitem [{\citenamefont {Goldschmidt}\ and\ \citenamefont
  {De~Dominicis}(1990)}]{goldschmidt1990replica}%
  \BibitemOpen
  \bibfield  {author} {\bibinfo {author} {\bibfnamefont {Y.~Y.}\ \bibnamefont
  {Goldschmidt}}\ and\ \bibinfo {author} {\bibfnamefont {C.}~\bibnamefont
  {De~Dominicis}},\ }\href@noop {} {\bibfield  {journal} {\bibinfo  {journal}
  {Physical Review B}\ }\textbf {\bibinfo {volume} {41}},\ \bibinfo {pages}
  {2184} (\bibinfo {year} {1990})}\BibitemShut {NoStop}%
\bibitem [{\citenamefont {Parisi}\ and\ \citenamefont
  {Tria}(2002)}]{parisi2002spin}%
  \BibitemOpen
  \bibfield  {author} {\bibinfo {author} {\bibfnamefont {G.}~\bibnamefont
  {Parisi}}\ and\ \bibinfo {author} {\bibfnamefont {F.}~\bibnamefont {Tria}},\
  }\href@noop {} {\bibfield  {journal} {\bibinfo  {journal} {The European
  Physical Journal B-Condensed Matter and Complex Systems}\ }\textbf {\bibinfo
  {volume} {30}},\ \bibinfo {pages} {533} (\bibinfo {year} {2002})}\BibitemShut
  {NoStop}%
\bibitem [{\citenamefont {Boschi}\ and\ \citenamefont
  {Parisi}(2020)}]{boschi2020free}%
  \BibitemOpen
  \bibfield  {author} {\bibinfo {author} {\bibfnamefont {G.}~\bibnamefont
  {Boschi}}\ and\ \bibinfo {author} {\bibfnamefont {G.}~\bibnamefont
  {Parisi}},\ }\href@noop {} {\bibfield  {journal} {\bibinfo  {journal} {arXiv
  preprint arXiv:2001.01966}\ } (\bibinfo {year} {2020})}\BibitemShut {NoStop}%
\bibitem [{\citenamefont {Mottishaw}(1987)}]{mottishaw1987replica}%
  \BibitemOpen
  \bibfield  {author} {\bibinfo {author} {\bibfnamefont {P.}~\bibnamefont
  {Mottishaw}},\ }\href@noop {} {\bibfield  {journal} {\bibinfo  {journal} {EPL
  (Europhysics Letters)}\ }\textbf {\bibinfo {volume} {4}},\ \bibinfo {pages}
  {333} (\bibinfo {year} {1987})}\BibitemShut {NoStop}%
\bibitem [{\citenamefont {Parisi}\ and\ \citenamefont
  {Rizzo}(2013)}]{parisi2013critical}%
  \BibitemOpen
  \bibfield  {author} {\bibinfo {author} {\bibfnamefont {G.}~\bibnamefont
  {Parisi}}\ and\ \bibinfo {author} {\bibfnamefont {T.}~\bibnamefont {Rizzo}},\
  }\href@noop {} {\bibfield  {journal} {\bibinfo  {journal} {Physical Review
  E}\ }\textbf {\bibinfo {volume} {87}},\ \bibinfo {pages} {012101} (\bibinfo
  {year} {2013})}\BibitemShut {NoStop}%
\bibitem [{\citenamefont {Rizzo}(2013)}]{rizzo2013replica}%
  \BibitemOpen
  \bibfield  {author} {\bibinfo {author} {\bibfnamefont {T.}~\bibnamefont
  {Rizzo}},\ }\href@noop {} {\bibfield  {journal} {\bibinfo  {journal}
  {Physical Review E}\ }\textbf {\bibinfo {volume} {88}},\ \bibinfo {pages}
  {032135} (\bibinfo {year} {2013})}\BibitemShut {NoStop}%
\bibitem [{\citenamefont {Parisi}\ \emph {et~al.}(2014)\citenamefont {Parisi},
  \citenamefont {Ricci-Tersenghi},\ and\ \citenamefont
  {Rizzo}}]{parisi2014diluted}%
  \BibitemOpen
  \bibfield  {author} {\bibinfo {author} {\bibfnamefont {G.}~\bibnamefont
  {Parisi}}, \bibinfo {author} {\bibfnamefont {F.}~\bibnamefont
  {Ricci-Tersenghi}},\ and\ \bibinfo {author} {\bibfnamefont {T.}~\bibnamefont
  {Rizzo}},\ }\href@noop {} {\bibfield  {journal} {\bibinfo  {journal} {Journal
  of Statistical Mechanics: Theory and Experiment}\ }\textbf {\bibinfo {volume}
  {2014}},\ \bibinfo {pages} {P04013} (\bibinfo {year} {2014})}\BibitemShut
  {NoStop}%
\bibitem [{\citenamefont {M\'{e}zard}\ and\ \citenamefont
  {Montanari}(2009)}]{mezard2009information}%
  \BibitemOpen
  \bibfield  {author} {\bibinfo {author} {\bibfnamefont {M.}~\bibnamefont
  {M\'{e}zard}}\ and\ \bibinfo {author} {\bibfnamefont {A.}~\bibnamefont
  {Montanari}},\ }\href@noop {} {\emph {\bibinfo {title} {{Information,
  Physics, and Computation}}}},\ Oxford Graduate Texts\ (\bibinfo  {publisher}
  {OUP Oxford},\ \bibinfo {year} {2009})\BibitemShut {NoStop}%
\bibitem [{\citenamefont {M{\'e}zard}\ \emph {et~al.}(1987)\citenamefont
  {M{\'e}zard}, \citenamefont {Parisi},\ and\ \citenamefont
  {Virasoro}}]{mezard1987spin}%
  \BibitemOpen
  \bibfield  {author} {\bibinfo {author} {\bibfnamefont {M.}~\bibnamefont
  {M{\'e}zard}}, \bibinfo {author} {\bibfnamefont {G.}~\bibnamefont {Parisi}},\
  and\ \bibinfo {author} {\bibfnamefont {M.}~\bibnamefont {Virasoro}},\
  }\href@noop {} {\emph {\bibinfo {title} {Spin Glass Theory and Beyond}}},\
  Lecture Notes in Physics Series\ (\bibinfo  {publisher} {World Scientific
  Publishing Company, Incorporated},\ \bibinfo {year} {1987})\BibitemShut
  {NoStop}%
\bibitem [{\citenamefont {M{\'e}zard}\ and\ \citenamefont
  {Zecchina}(2002)}]{mezard2002random}%
  \BibitemOpen
  \bibfield  {author} {\bibinfo {author} {\bibfnamefont {M.}~\bibnamefont
  {M{\'e}zard}}\ and\ \bibinfo {author} {\bibfnamefont {R.}~\bibnamefont
  {Zecchina}},\ }\href@noop {} {\bibfield  {journal} {\bibinfo  {journal}
  {Physical Review E}\ }\textbf {\bibinfo {volume} {66}},\ \bibinfo {pages}
  {056126} (\bibinfo {year} {2002})}\BibitemShut {NoStop}%
\bibitem [{\citenamefont {Morone}\ \emph {et~al.}(2014)\citenamefont {Morone},
  \citenamefont {Parisi},\ and\ \citenamefont
  {Ricci-Tersenghi}}]{morone2014large}%
  \BibitemOpen
  \bibfield  {author} {\bibinfo {author} {\bibfnamefont {F.}~\bibnamefont
  {Morone}}, \bibinfo {author} {\bibfnamefont {G.}~\bibnamefont {Parisi}},\
  and\ \bibinfo {author} {\bibfnamefont {F.}~\bibnamefont {Ricci-Tersenghi}},\
  }\href@noop {} {\bibfield  {journal} {\bibinfo  {journal} {Physical Review
  B}\ }\textbf {\bibinfo {volume} {89}},\ \bibinfo {pages} {214202} (\bibinfo
  {year} {2014})}\BibitemShut {NoStop}%
\bibitem [{\citenamefont {Krzakala}\ \emph {et~al.}(2010)\citenamefont
  {Krzakala}, \citenamefont {Ricci-Tersenghi},\ and\ \citenamefont
  {Zdeborov{\'a}}}]{krzakala2010elusive}%
  \BibitemOpen
  \bibfield  {author} {\bibinfo {author} {\bibfnamefont {F.}~\bibnamefont
  {Krzakala}}, \bibinfo {author} {\bibfnamefont {F.}~\bibnamefont
  {Ricci-Tersenghi}},\ and\ \bibinfo {author} {\bibfnamefont {L.}~\bibnamefont
  {Zdeborov{\'a}}},\ }\href@noop {} {\bibfield  {journal} {\bibinfo  {journal}
  {Physical review letters}\ }\textbf {\bibinfo {volume} {104}},\ \bibinfo
  {pages} {207208} (\bibinfo {year} {2010})}\BibitemShut {NoStop}%
\bibitem [{\citenamefont {Chatterjee}(2015)}]{chatterjee2015absence}%
  \BibitemOpen
  \bibfield  {author} {\bibinfo {author} {\bibfnamefont {S.}~\bibnamefont
  {Chatterjee}},\ }\href@noop {} {\bibfield  {journal} {\bibinfo  {journal}
  {Communications in Mathematical Physics}\ }\textbf {\bibinfo {volume}
  {337}},\ \bibinfo {pages} {93} (\bibinfo {year} {2015})}\BibitemShut
  {NoStop}%
\bibitem [{\citenamefont {Perugini}\ and\ \citenamefont
  {Ricci-Tersenghi}(2018)}]{perugini2018improved}%
  \BibitemOpen
  \bibfield  {author} {\bibinfo {author} {\bibfnamefont {G.}~\bibnamefont
  {Perugini}}\ and\ \bibinfo {author} {\bibfnamefont {F.}~\bibnamefont
  {Ricci-Tersenghi}},\ }\href@noop {} {\bibfield  {journal} {\bibinfo
  {journal} {Physical Review E}\ }\textbf {\bibinfo {volume} {97}},\ \bibinfo
  {pages} {012152} (\bibinfo {year} {2018})}\BibitemShut {NoStop}%
\bibitem [{\citenamefont {Angelini}\ \emph {et~al.}(2020)\citenamefont
  {Angelini}, \citenamefont {Lucibello}, \citenamefont {Parisi}, \citenamefont
  {Ricci-Tersenghi},\ and\ \citenamefont {Rizzo}}]{angelini2020loop}%
  \BibitemOpen
  \bibfield  {author} {\bibinfo {author} {\bibfnamefont {M.~C.}\ \bibnamefont
  {Angelini}}, \bibinfo {author} {\bibfnamefont {C.}~\bibnamefont {Lucibello}},
  \bibinfo {author} {\bibfnamefont {G.}~\bibnamefont {Parisi}}, \bibinfo
  {author} {\bibfnamefont {F.}~\bibnamefont {Ricci-Tersenghi}},\ and\ \bibinfo
  {author} {\bibfnamefont {T.}~\bibnamefont {Rizzo}},\ }\href@noop {}
  {\bibfield  {journal} {\bibinfo  {journal} {Proceedings of the National
  Academy of Sciences}\ }\textbf {\bibinfo {volume} {117}},\ \bibinfo {pages}
  {2268} (\bibinfo {year} {2020})}\BibitemShut {NoStop}%
\bibitem [{\citenamefont {Angelini}\ \emph {et~al.}(2022)\citenamefont
  {Angelini}, \citenamefont {Lucibello}, \citenamefont {Parisi}, \citenamefont
  {Perrupato}, \citenamefont {Ricci-Tersenghi},\ and\ \citenamefont
  {Rizzo}}]{angelini2022unexpected}%
  \BibitemOpen
  \bibfield  {author} {\bibinfo {author} {\bibfnamefont {M.~C.}\ \bibnamefont
  {Angelini}}, \bibinfo {author} {\bibfnamefont {C.}~\bibnamefont {Lucibello}},
  \bibinfo {author} {\bibfnamefont {G.}~\bibnamefont {Parisi}}, \bibinfo
  {author} {\bibfnamefont {G.}~\bibnamefont {Perrupato}}, \bibinfo {author}
  {\bibfnamefont {F.}~\bibnamefont {Ricci-Tersenghi}},\ and\ \bibinfo {author}
  {\bibfnamefont {T.}~\bibnamefont {Rizzo}},\ }\href@noop {} {\bibfield
  {journal} {\bibinfo  {journal} {Physical Review Letters}\ }\textbf {\bibinfo
  {volume} {128}},\ \bibinfo {pages} {075702} (\bibinfo {year}
  {2022})}\BibitemShut {NoStop}%
\bibitem [{\citenamefont {Le~Doussal}\ \emph {et~al.}(2009)\citenamefont
  {Le~Doussal}, \citenamefont {Middleton},\ and\ \citenamefont
  {Wiese}}]{le2009statistics}%
  \BibitemOpen
  \bibfield  {author} {\bibinfo {author} {\bibfnamefont {P.}~\bibnamefont
  {Le~Doussal}}, \bibinfo {author} {\bibfnamefont {A.~A.}\ \bibnamefont
  {Middleton}},\ and\ \bibinfo {author} {\bibfnamefont {K.~J.}\ \bibnamefont
  {Wiese}},\ }\href@noop {} {\bibfield  {journal} {\bibinfo  {journal}
  {Physical Review E}\ }\textbf {\bibinfo {volume} {79}},\ \bibinfo {pages}
  {050101(R)} (\bibinfo {year} {2009})}\BibitemShut {NoStop}%
\bibitem [{\citenamefont {Le~Doussal}\ \emph {et~al.}(2010)\citenamefont
  {Le~Doussal}, \citenamefont {M{\"u}ller},\ and\ \citenamefont
  {Wiese}}]{le2010avalanches}%
  \BibitemOpen
  \bibfield  {author} {\bibinfo {author} {\bibfnamefont {P.}~\bibnamefont
  {Le~Doussal}}, \bibinfo {author} {\bibfnamefont {M.}~\bibnamefont
  {M{\"u}ller}},\ and\ \bibinfo {author} {\bibfnamefont {K.~J.}\ \bibnamefont
  {Wiese}},\ }\href@noop {} {\bibfield  {journal} {\bibinfo  {journal} {EPL
  (Europhysics Letters)}\ }\textbf {\bibinfo {volume} {91}},\ \bibinfo {pages}
  {57004} (\bibinfo {year} {2010})}\BibitemShut {NoStop}%
\bibitem [{\citenamefont {Tarjus}\ \emph {et~al.}(2013)\citenamefont {Tarjus},
  \citenamefont {Baczyk},\ and\ \citenamefont
  {Tissier}}]{tarjus2013avalanches}%
  \BibitemOpen
  \bibfield  {author} {\bibinfo {author} {\bibfnamefont {G.}~\bibnamefont
  {Tarjus}}, \bibinfo {author} {\bibfnamefont {M.}~\bibnamefont {Baczyk}},\
  and\ \bibinfo {author} {\bibfnamefont {M.}~\bibnamefont {Tissier}},\
  }\href@noop {} {\bibfield  {journal} {\bibinfo  {journal} {Physical review
  letters}\ }\textbf {\bibinfo {volume} {110}},\ \bibinfo {pages} {135703}
  (\bibinfo {year} {2013})}\BibitemShut {NoStop}%
\bibitem [{\citenamefont {Altieri}\ \emph {et~al.}(2017)\citenamefont
  {Altieri}, \citenamefont {Angelini}, \citenamefont {Lucibello}, \citenamefont
  {Parisi}, \citenamefont {Ricci-Tersenghi},\ and\ \citenamefont
  {Rizzo}}]{altieri2017loop}%
  \BibitemOpen
  \bibfield  {author} {\bibinfo {author} {\bibfnamefont {A.}~\bibnamefont
  {Altieri}}, \bibinfo {author} {\bibfnamefont {M.~C.}\ \bibnamefont
  {Angelini}}, \bibinfo {author} {\bibfnamefont {C.}~\bibnamefont {Lucibello}},
  \bibinfo {author} {\bibfnamefont {G.}~\bibnamefont {Parisi}}, \bibinfo
  {author} {\bibfnamefont {F.}~\bibnamefont {Ricci-Tersenghi}},\ and\ \bibinfo
  {author} {\bibfnamefont {T.}~\bibnamefont {Rizzo}},\ }\href@noop {}
  {\bibfield  {journal} {\bibinfo  {journal} {Journal of Statistical Mechanics:
  Theory and Experiment}\ }\textbf {\bibinfo {volume} {2017}},\ \bibinfo
  {pages} {113303} (\bibinfo {year} {2017})}\BibitemShut {NoStop}%
\bibitem [{\citenamefont {Bray}\ and\ \citenamefont
  {Roberts}(1980)}]{bray1980renormalisation}%
  \BibitemOpen
  \bibfield  {author} {\bibinfo {author} {\bibfnamefont {A.}~\bibnamefont
  {Bray}}\ and\ \bibinfo {author} {\bibfnamefont {S.}~\bibnamefont {Roberts}},\
  }\href@noop {} {\bibfield  {journal} {\bibinfo  {journal} {J.\ Phys.\ C}\
  }\textbf {\bibinfo {volume} {13}},\ \bibinfo {pages} {5405} (\bibinfo {year}
  {1980})}\BibitemShut {NoStop}%
\bibitem [{\citenamefont {Moore}\ and\ \citenamefont
  {Bray}(2011)}]{moore2011disappearance}%
  \BibitemOpen
  \bibfield  {author} {\bibinfo {author} {\bibfnamefont {M.~A.}\ \bibnamefont
  {Moore}}\ and\ \bibinfo {author} {\bibfnamefont {A.~J.}\ \bibnamefont
  {Bray}},\ }\href@noop {} {\bibfield  {journal} {\bibinfo  {journal} {Phys.\
  Rev.\ B}\ }\textbf {\bibinfo {volume} {83}},\ \bibinfo {pages} {224408}
  (\bibinfo {year} {2011})}\BibitemShut {NoStop}%
\bibitem [{\citenamefont {Parisi}\ and\ \citenamefont
  {Temesv{\'a}ri}(2012)}]{parisi2012replica}%
  \BibitemOpen
  \bibfield  {author} {\bibinfo {author} {\bibfnamefont {G.}~\bibnamefont
  {Parisi}}\ and\ \bibinfo {author} {\bibfnamefont {T.}~\bibnamefont
  {Temesv{\'a}ri}},\ }\href@noop {} {\bibfield  {journal} {\bibinfo  {journal}
  {Nucl.\ Phys.\ B}\ }\textbf {\bibinfo {volume} {858}},\ \bibinfo {pages}
  {293} (\bibinfo {year} {2012})}\BibitemShut {NoStop}%
\bibitem [{\citenamefont {Angelini}\ and\ \citenamefont
  {Biroli}(2015)}]{angelini2015spin}%
  \BibitemOpen
  \bibfield  {author} {\bibinfo {author} {\bibfnamefont {M.~C.}\ \bibnamefont
  {Angelini}}\ and\ \bibinfo {author} {\bibfnamefont {G.}~\bibnamefont
  {Biroli}},\ }\href@noop {} {\bibfield  {journal} {\bibinfo  {journal}
  {Physical review letters}\ }\textbf {\bibinfo {volume} {114}},\ \bibinfo
  {pages} {095701} (\bibinfo {year} {2015})}\BibitemShut {NoStop}%
\bibitem [{\citenamefont {Angelini}\ and\ \citenamefont
  {Biroli}(2017)}]{angelini2017real}%
  \BibitemOpen
  \bibfield  {author} {\bibinfo {author} {\bibfnamefont {M.~C.}\ \bibnamefont
  {Angelini}}\ and\ \bibinfo {author} {\bibfnamefont {G.}~\bibnamefont
  {Biroli}},\ }\href@noop {} {\bibfield  {journal} {\bibinfo  {journal}
  {Journal of Statistical Physics}\ }\textbf {\bibinfo {volume} {167}},\
  \bibinfo {pages} {476} (\bibinfo {year} {2017})}\BibitemShut {NoStop}%
\bibitem [{\citenamefont {Urbani}(2022)}]{urbani2022field}%
  \BibitemOpen
  \bibfield  {author} {\bibinfo {author} {\bibfnamefont {P.}~\bibnamefont
  {Urbani}},\ }\href@noop {} {\bibfield  {journal} {\bibinfo  {journal} {arXiv
  preprint arXiv:2203.01899}\ } (\bibinfo {year} {2022})}\BibitemShut {NoStop}%
\bibitem [{Note1()}]{Note1}%
  \BibitemOpen
  \bibinfo {note} {Note that Eq.~\protect \textup {\hbox {\mathsurround \z@
  \protect \normalfont (\ignorespaces \ref {eq:leadingMcorr}\unskip
  \@@italiccorr )}} takes an analogous form to that of the finite size
  corrections to disorder models on sparse graphs \cite
  {ferrari2013finite,parisi2020random}}\BibitemShut {NoStop}%
\bibitem [{\citenamefont {Franz}\ and\ \citenamefont
  {Parisi}(2000{\natexlab{a}})}]{Franz2000}%
  \BibitemOpen
  \bibfield  {author} {\bibinfo {author} {\bibfnamefont {S.}~\bibnamefont
  {Franz}}\ and\ \bibinfo {author} {\bibfnamefont {G.}~\bibnamefont {Parisi}},\
  }\href {https://doi.org/10.1007/s100510070037} {\bibfield  {journal}
  {\bibinfo  {journal} {Eur. Phys. J. B}\ }\textbf {\bibinfo {volume} {18}},\
  \bibinfo {pages} {485} (\bibinfo {year} {2000}{\natexlab{a}})}\BibitemShut
  {NoStop}%
\bibitem [{\citenamefont {Vannimenus}\ \emph {et~al.}(1981)\citenamefont
  {Vannimenus}, \citenamefont {Toulouse},\ and\ \citenamefont
  {Parisi}}]{vannimenus1981study}%
  \BibitemOpen
  \bibfield  {author} {\bibinfo {author} {\bibfnamefont {J.}~\bibnamefont
  {Vannimenus}}, \bibinfo {author} {\bibfnamefont {G.}~\bibnamefont
  {Toulouse}},\ and\ \bibinfo {author} {\bibfnamefont {G.}~\bibnamefont
  {Parisi}},\ }\href@noop {} {\bibfield  {journal} {\bibinfo  {journal}
  {Journal de Physique}\ }\textbf {\bibinfo {volume} {42}},\ \bibinfo {pages}
  {565} (\bibinfo {year} {1981})}\BibitemShut {NoStop}%
\bibitem [{\citenamefont {Parisi}\ and\ \citenamefont
  {Toulouse}(1980)}]{parisi1980simple}%
  \BibitemOpen
  \bibfield  {author} {\bibinfo {author} {\bibfnamefont {G.}~\bibnamefont
  {Parisi}}\ and\ \bibinfo {author} {\bibfnamefont {G.}~\bibnamefont
  {Toulouse}},\ }\href@noop {} {\bibfield  {journal} {\bibinfo  {journal}
  {Journal de Physique Lettres}\ }\textbf {\bibinfo {volume} {41}},\ \bibinfo
  {pages} {361} (\bibinfo {year} {1980})}\BibitemShut {NoStop}%
\bibitem [{\citenamefont {Franz}\ and\ \citenamefont
  {Parisi}(2000{\natexlab{b}})}]{franz2000non}%
  \BibitemOpen
  \bibfield  {author} {\bibinfo {author} {\bibfnamefont {S.}~\bibnamefont
  {Franz}}\ and\ \bibinfo {author} {\bibfnamefont {G.}~\bibnamefont {Parisi}},\
  }\href@noop {} {\bibfield  {journal} {\bibinfo  {journal} {The European
  Physical Journal B-Condensed Matter and Complex Systems}\ }\textbf {\bibinfo
  {volume} {18}},\ \bibinfo {pages} {485} (\bibinfo {year}
  {2000}{\natexlab{b}})}\BibitemShut {NoStop}%
\bibitem [{\citenamefont {Semerjian}(2008)}]{semerjian2008freezing}%
  \BibitemOpen
  \bibfield  {author} {\bibinfo {author} {\bibfnamefont {G.}~\bibnamefont
  {Semerjian}},\ }\href@noop {} {\bibfield  {journal} {\bibinfo  {journal}
  {Journal of Statistical Physics}\ }\textbf {\bibinfo {volume} {130}},\
  \bibinfo {pages} {251} (\bibinfo {year} {2008})}\BibitemShut {NoStop}%
\bibitem [{\citenamefont {Panchenko}\ and\ \citenamefont
  {Talagrand}(2004)}]{panchenko2004bounds}%
  \BibitemOpen
  \bibfield  {author} {\bibinfo {author} {\bibfnamefont {D.}~\bibnamefont
  {Panchenko}}\ and\ \bibinfo {author} {\bibfnamefont {M.}~\bibnamefont
  {Talagrand}},\ }\href@noop {} {\bibfield  {journal} {\bibinfo  {journal}
  {Probability Theory and Related Fields}\ }\textbf {\bibinfo {volume} {130}},\
  \bibinfo {pages} {319} (\bibinfo {year} {2004})}\BibitemShut {NoStop}%
\bibitem [{\citenamefont {Franz}\ and\ \citenamefont
  {Leone}(2003)}]{franz2003replica}%
  \BibitemOpen
  \bibfield  {author} {\bibinfo {author} {\bibfnamefont {S.}~\bibnamefont
  {Franz}}\ and\ \bibinfo {author} {\bibfnamefont {M.}~\bibnamefont {Leone}},\
  }\href@noop {} {\bibfield  {journal} {\bibinfo  {journal} {Journal of
  Statistical Physics}\ }\textbf {\bibinfo {volume} {111}},\ \bibinfo {pages}
  {535} (\bibinfo {year} {2003})}\BibitemShut {NoStop}%
\bibitem [{\citenamefont {Guerra}(2003)}]{guerra2003broken}%
  \BibitemOpen
  \bibfield  {author} {\bibinfo {author} {\bibfnamefont {F.}~\bibnamefont
  {Guerra}},\ }\href@noop {} {\bibfield  {journal} {\bibinfo  {journal}
  {Communications in mathematical physics}\ }\textbf {\bibinfo {volume}
  {233}},\ \bibinfo {pages} {1} (\bibinfo {year} {2003})}\BibitemShut {NoStop}%
\bibitem [{\citenamefont {Montanari}\ \emph {et~al.}(2004)\citenamefont
  {Montanari}, \citenamefont {Parisi},\ and\ \citenamefont
  {Ricci-Tersenghi}}]{montanari2004instability}%
  \BibitemOpen
  \bibfield  {author} {\bibinfo {author} {\bibfnamefont {A.}~\bibnamefont
  {Montanari}}, \bibinfo {author} {\bibfnamefont {G.}~\bibnamefont {Parisi}},\
  and\ \bibinfo {author} {\bibfnamefont {F.}~\bibnamefont {Ricci-Tersenghi}},\
  }\href@noop {} {\bibfield  {journal} {\bibinfo  {journal} {Journal of Physics
  A: Mathematical and General}\ }\textbf {\bibinfo {volume} {37}},\ \bibinfo
  {pages} {2073} (\bibinfo {year} {2004})}\BibitemShut {NoStop}%
\bibitem [{\citenamefont {Marinari}\ \emph {et~al.}(2000)\citenamefont
  {Marinari}, \citenamefont {Parisi}, \citenamefont {Ricci-Tersenghi},
  \citenamefont {Ruiz-Lorenzo},\ and\ \citenamefont
  {Zuliani}}]{marinari2000replica}%
  \BibitemOpen
  \bibfield  {author} {\bibinfo {author} {\bibfnamefont {E.}~\bibnamefont
  {Marinari}}, \bibinfo {author} {\bibfnamefont {G.}~\bibnamefont {Parisi}},
  \bibinfo {author} {\bibfnamefont {F.}~\bibnamefont {Ricci-Tersenghi}},
  \bibinfo {author} {\bibfnamefont {J.~J.}\ \bibnamefont {Ruiz-Lorenzo}},\ and\
  \bibinfo {author} {\bibfnamefont {F.}~\bibnamefont {Zuliani}},\ }\href@noop
  {} {\bibfield  {journal} {\bibinfo  {journal} {Journal of Statistical
  Physics}\ }\textbf {\bibinfo {volume} {98}},\ \bibinfo {pages} {973}
  (\bibinfo {year} {2000})}\BibitemShut {NoStop}%
\bibitem [{\citenamefont {Angelini}\ \emph {et~al.}(2019)\citenamefont
  {Angelini}, \citenamefont {Lucibello}, \citenamefont {Parisi}, \citenamefont
  {Ricci-Tersenghi},\ and\ \citenamefont {Rizzo}}]{angelini2019new}%
  \BibitemOpen
  \bibfield  {author} {\bibinfo {author} {\bibfnamefont {M.~C.}\ \bibnamefont
  {Angelini}}, \bibinfo {author} {\bibfnamefont {C.}~\bibnamefont {Lucibello}},
  \bibinfo {author} {\bibfnamefont {G.}~\bibnamefont {Parisi}}, \bibinfo
  {author} {\bibfnamefont {F.}~\bibnamefont {Ricci-Tersenghi}},\ and\ \bibinfo
  {author} {\bibfnamefont {T.}~\bibnamefont {Rizzo}},\ }\href@noop {}
  {\bibfield  {journal} {\bibinfo  {journal} {arXiv preprint arXiv:1906.04437}\
  } (\bibinfo {year} {2019})}\BibitemShut {NoStop}%
\bibitem [{\citenamefont {Stauffer}\ and\ \citenamefont
  {Aharony}(2018)}]{stauffer2018introduction}%
  \BibitemOpen
  \bibfield  {author} {\bibinfo {author} {\bibfnamefont {D.}~\bibnamefont
  {Stauffer}}\ and\ \bibinfo {author} {\bibfnamefont {A.}~\bibnamefont
  {Aharony}},\ }\href@noop {} {\emph {\bibinfo {title} {Introduction to
  percolation theory}}}\ (\bibinfo  {publisher} {CRC press},\ \bibinfo {year}
  {2018})\BibitemShut {NoStop}%
\bibitem [{\citenamefont {Ferrari}\ \emph {et~al.}(2013)\citenamefont
  {Ferrari}, \citenamefont {Lucibello}, \citenamefont {Morone}, \citenamefont
  {Parisi}, \citenamefont {Ricci-Tersenghi},\ and\ \citenamefont
  {Rizzo}}]{ferrari2013finite}%
  \BibitemOpen
  \bibfield  {author} {\bibinfo {author} {\bibfnamefont {U.}~\bibnamefont
  {Ferrari}}, \bibinfo {author} {\bibfnamefont {C.}~\bibnamefont {Lucibello}},
  \bibinfo {author} {\bibfnamefont {F.}~\bibnamefont {Morone}}, \bibinfo
  {author} {\bibfnamefont {G.}~\bibnamefont {Parisi}}, \bibinfo {author}
  {\bibfnamefont {F.}~\bibnamefont {Ricci-Tersenghi}},\ and\ \bibinfo {author}
  {\bibfnamefont {T.}~\bibnamefont {Rizzo}},\ }\href@noop {} {\bibfield
  {journal} {\bibinfo  {journal} {Physical Review B}\ }\textbf {\bibinfo
  {volume} {88}},\ \bibinfo {pages} {184201} (\bibinfo {year}
  {2013})}\BibitemShut {NoStop}%
\bibitem [{\citenamefont {Parisi}\ \emph {et~al.}(2020)\citenamefont {Parisi},
  \citenamefont {Perrupato},\ and\ \citenamefont {Sicuro}}]{parisi2020random}%
  \BibitemOpen
  \bibfield  {author} {\bibinfo {author} {\bibfnamefont {G.}~\bibnamefont
  {Parisi}}, \bibinfo {author} {\bibfnamefont {G.}~\bibnamefont {Perrupato}},\
  and\ \bibinfo {author} {\bibfnamefont {G.}~\bibnamefont {Sicuro}},\
  }\href@noop {} {\bibfield  {journal} {\bibinfo  {journal} {Journal of
  Statistical Mechanics: Theory and Experiment}\ }\textbf {\bibinfo {volume}
  {2020}},\ \bibinfo {pages} {033301} (\bibinfo {year} {2020})}\BibitemShut
  {NoStop}%
\end{thebibliography}%

\end{document}